\documentclass[useAMS,usenatbib,,usegraphicx,onecolumn]{mn2e}
\usepackage{txfonts}
\makeatletter

\DeclareMathAlphabet{\mathpzc}{OT1}{pzc}{m}{it}
\newcommand\I{\mathrm i}
\newcommand\D{\mathrm d}
\newcommand\E{\mathrm e}
\newcommand\f{\bar f}
\newcommand\g{\bar g}
\newcommand\w{\bar w}
\newcommand\z{\bar z}
\newcommand\ab{\bar a}
\newcommand\lb{\bar l}
\newcommand\J{\mathcal J}
\newcommand\ds{\mathpzc d}
\newcommand\qr{\mathpzc q}
\newcommand\ee{\mathpzc e}
\newcommand\fz{\mathpzc f}
\newcommand\gz{\mathpzc g}
\newcommand\rdash{\mbox{${r\mkern-8mu\mbox{--}}$}}
\newcommand\plot{\centering\includegraphics[width=0.65\hsize]}

\eqsecnum

\pagerange{\pageref{start}--\pageref{finish}}\pubyear{2004}
\renewcommand{\today}{15th Dec 2004}
\date{doi:10.1111/j.1365-2966.2004.08581.x}
\journal{\sl to appear in
\it Monthly Notices of the Royal Astronomical Society;
\sc pp.\pageref{start}--\pageref{finish} (\today: v1.4.1)}

\title[SYMMETRY OF PERTURBATION LENSING POTENTIALS]
{Gravitational lens under perturbations:
Symmetry of perturbing potentials with invariant caustics}
\author[J.~An]{Jin H. An\thanks{E-mail: jin@ast.cam.ac.uk}\\
Institute of Astronomy, University of Cambridge,
Madingley Road, Cambridge CB3 0HA, UK}

\begin{document}
\label{start}
\maketitle

\begin{abstract}
When the gravitational lensing potential can be approximated by that
of a circularly symmetric system affected by weak perturbations, it is
found that the shape of the resulting (tangential) caustics is
entirely specified by the local azimuthal behaviour of the affecting
perturbations. This provides a common mathematical groundwork for
understanding problems such as the close-wide ($d\leftrightarrow
d^{-1}$) separation degeneracy of binary lens microlensing
lightcurves and the shear-ellipticity degeneracy of quadruple image
lens modelling.
\end{abstract}


\section{Introduction}

The recent announcement by \citet{Bo04} of the most convincing planetary
microlensing event \citep{MP91} to date is one of the greatest
success stories of the nearly two-decade old promise of microlensing
\citep{Pa86}. However, despite such successes, the study of
microlensing is still hampered by various hurdles, one of the most
persistent being the problem of degenerate lightcurves
\citep[e.g.,][]{Do99a,Af00} -- that is, several distinct physical systems
can be used to model observed lightcurves.

For the purpose of formal discussion, the degeneracy of microlensing
lightcurves may be categorized into two separate but
related phenomena. The first, which may be termed as
`external' degeneracy, is caused by imperfect observations yielding
highly correlated measurements of the parameters that control the
observed lightcurves. For instance, in most highly blended events,
without extremely good coverage of certain wing portions of the
lightcurve, one only expects to measure the ratio of the
time-scale to the peak magnification (or the product of the time-scale and
the source flux) but not the parameters individually \citep[c.f.,][]{Go96}.
On the other hand, the second type of degeneracy, which will henceforth
be referred to as `internal' degeneracy, is due to the fact that the
number of independent lightcurve-controlling parameters is smaller
than the number of parameters that are used to describe the underlying
lens systems. In other words, even with perfectly specified
lightcurves (consequently, perfectly determined lightcurve
parameters), the underlying lens system may not be uniquely recovered.
This internal degeneracy of microlensing lightcurves can be
further divided into two broad classes; the `intrinsic' degeneracy
among magnification structures from different systems, and different
lens systems tracing identical or intrinsically degenerate paths in
the given magnification structure (hereafter the `extrinsic'
degeneracy). The most widely acknowledged degeneracy of microlensing
lightcurves, that is, the `time-scale degeneracy' relating to the mass
of the lens, the distances to the source and the lens, and the
relative motions between the observer, the source, and the lens, is an
example of the extrinsic degeneracy. Other examples of the extrinsic
degeneracy are the discrete parallax degeneracies
\citep*[e.g.,][]{SMP03,Go04}. However, it is the intrinsic degeneracy
that is of the most theoretical interest as it can be studied through
the analysis of the lens equation as a whole without specified
reference to the path of the system through the magnification
structure or the observational conditions. Furthermore, understanding
of the intrinsic degeneracy can greatly facilitate the
identification of various extrinsic and external degeneracies. The most
obvious example of an intrinsic degeneracy is the azimuthal
symmetry as well as the radial similarity of the lensing magnification
of the point mass lens. The well-known result of the `inverse-square-root'
behaviour of magnification `inside' fold caustics
\citep[e.g.,][]{GP02} is a less obvious example of a \emph{local}
intrinsic degeneracy.

It is notable that the intrinsic degeneracy is intimately
related to the symmetry of the lens equation.
In general, if (the potential of) the system possesses a certain
symmetry, one can expect that there should exist an intrinsic
degeneracy of magnification related to it. Note that the magnification
is basically the second derivative of the potential. This also means
that the study of the intrinsic degeneracy beyond the incidental case studies
can be greatly systematized by recognizing the associated
symmetry structure of the lens system. However, beyond
the obvious example of the circularly symmetric lens, studies of
degeneracies rooted in symmetries have been minimal, at best.
This paper is one of the first attempts to understand certain intrinsic
degeneracies of microlensing by searching for a symmetry in the lens
system. It is found that this approach proves itself
by providing a unified scheme of understanding the problem of
a certain well-known microlensing degeneracy and
a seemingly remote problem of lens modelling of a quadruple images.

The primary focus of this manuscript is a point mass lens (and a more
general circularly symmetric lens in later sections) that is under the
external influence of certain perturbations. It should be noted not only
that the symmetry in such systems can be analysed with relative ease,
but also that they can be applied to various realistic systems.
In Section~\ref{sec_scl}, the basic properties of the point mass
lens are recapitulated, and in Section~\ref{sec_per},
a perturbative approach to gravitational lensing is developed. In
particular, in section~\ref{ssec_ccc}, the perturbative approach is
used to derive an approximate expression for caustics and
critical curves. Following this, in Section~\ref{sec_isp},
it is shown that there exists a certain set of perturbations that yields
invariant caustics, which may be seen as the main finding of this
monograph. This finding is applied to interpret one of the well-known
microlensing degeneracy problems in the next Section~\ref{sec_egs}.
The restrictions imposed on the studied lens system are subsequently
relaxed in Section~\ref{sec_dis}. For example,
the discussion is extended from a point mass lens under perturbations to
a general circularly symmetric lens system under perturbations
in Section~\ref{ssec_csl}, and from external perturbations to
perturbations associated with the system itself in Section~\ref{ssec_scp}.
These two generalizations lay the basis for a study
of the shear and ellipticity in lens modelling later in the same section.

In this paper, the lens equation formulated in terms of complex numbers
\citep{BK75,Wi90,WM95} is extensively used. In Appendix~\ref{asec_caf},
most of the mathematical terminologies regarding complex analysis found
in this manuscript are summarized, and in Appendix~\ref{asec_ceq}, the
basic gravitational lens theory is redeveloped using complex-number
notation. The treatments given there are minimal and
deliberately casual. More in-depth treatments of complex analysis
including rigorous definitions and proofs can be found in any standard
text of complex analysis \citep[e.g.,][]{Ah79}
or of mathematical methods \citep[e.g.,][]{AW00}. Interested
readers may also consult references regarding potential theory in
a plane.

\section{Schwarzschild lens}
\label{sec_scl}

When the lens system can be approximated by a Schwarzschild metric,
one can write the lens equation in complex notation (see
Appendix~\ref{asec_ceq}) as
\begin{equation}
\label{le0}
\zeta_s=z-\frac{1}{\z},
\end{equation}
which relates the apparent angular position of the lensed image $z$ to
the (would-be) angular position of the source $\zeta_s$ in the absence
of the lens. Note that, throughout this paper, the complex conjugate
is represented with an overline (or upper bar) so that $\z$ is the complex
conjugate of $z$. Here, the lens is located at the coordinate origin,
and all the angular measurements are made in units of the angular
Einstein ring radius,
\begin{equation}
\label{enr}
\theta_\mathrm{E}=\left(\frac{2R_\mathrm{Sch}}{D_\mathrm{rel}}\right)^{1/2},
\end{equation}
where $R_\mathrm{Sch}\equiv2Gmc^{-2}$ is Schwarzschild radius of the
lens mass $m$, and $D_\mathrm{rel}\equiv D_\mathrm{LS}^{-1}
D_\mathrm{L} D_\mathrm{S}$ is the relative parallax
distance\footnote{If $D_\mathrm{LS}=D_\mathrm{S}-D_\mathrm{L}$, then
$D_\mathrm{rel}^{-1}=D_\mathrm{L}^{-1}-D_\mathrm{S}^{-1}$.}, and
$D_\mathrm{S}$ and $D_\mathrm{L}$ are the distances to the source and
to the lens respectively, while $D_\mathrm{LS}$ is the distance to the
source from the lens.

It is relatively straightforward, if not entirely trivial, to solve
the lens equation~(\ref{le0}) to find the image positions $z$ for a
given source position $\zeta_s$. However, one can discover a few
interesting properties of the lens equation~(\ref{le0})
before actually solving it. For example, from the symmetry
of the equation, one can immediately find that, if $z_1$ is
an image position for a given source position, $z_2=-\z_1^{-1}$ should
also be an image position for the same source position (also note
that $\z_2^{-1}=-z_1$). In fact, it is easy to show that
equation~(\ref{le0}) always allows only two images for any given source
position $\zeta_s$ unless $\zeta_s=0$, in which case the
equation~(\ref{le0}) is reduced to the equation of a unit circle,
$|z|=1$. That is, the image becomes a ring (which is also naturally
expected from the intrinsic circular symmetry of the system) with its
radius given by equation~(\ref{enr}).
This ringed image is sometimes referred to as `Einstein ring.'

One important quantitative measure in gravitational lensing is
the magnification factor of lensed images. In a purely mathematical
sense, the lens equation defines a mapping from the image position to
the source position, both of which are vectors in two dimensional
spaces. The local -- differential -- behaviour of this mapping therefore
can be studied from its Jacobian, which is basically a linear
transformation that approximates the lens mapping locally. It follows
naturally that the Jacobian determinant of any given lens mapping is
the inverse magnification factor of the lensed image in the limit of a
point source. If the lens mapping is given by the lens
equation~(\ref{le0}), one can easily show that the associated Jacobian
determinant is that
\begin{equation}
\label{sjd}
\J_s=
\upartial_z\zeta_s\upartial_{\z}\bar\zeta_s-
\upartial_{\z}\zeta_s\upartial_z\bar\zeta_s=
|\upartial_z\zeta_s|^2-|\upartial_{\z}\zeta_s|^2=
1-\frac{1}{|z|^4}.
\end{equation}
Here, $\upartial_z\equiv(\partial/\partial z)$ is simplified
notation for the partial derivative operator. Equation~(\ref{sjd}) gives
the inverse magnification for a single image. If one wants to
find the total magnification, accounting for all images for a given
source position, one needs to find the sum of the inverse absolute
values of the Jacobian determinants corresponding to all the images. It is
also notable that the Jacobian determinant given by
equation~(\ref{sjd}) vanishes when $|z|=1$ (consequently $\zeta_s=0$),
which implies that the magnification for this image is formally
infinite -- the lensed image of a `0'-dimensional point source is a
`1'-dimensional ring. In general, the loci in the image space defined
by $\J=0$ are referred to as critical curves, and the image of the
critical curves under the lens mapping in the source plane as
caustics. Thus, for the point mass lens case, the caustic is a point
($\zeta_s=0$), and the critical curve is a circle ($|z|=1$).

\section{slowly-varying small null-convergent perturbation}
\label{sec_per}

Let us consider the situation in which a point mass lens is subject to
some null-convergent -- i.e., the continuous surface lens mass density
is zero except finite number of isolated mass points --
perturbations. Then, the perturbation part of the lensing potential
$\psi$ satisfies the two-dimensional Laplace equation
$\nabla^2\psi=0$.
Since the real and the imaginary parts of any complex analytic
function are harmonic, there exists a complex analytic function
$\psi_c(z)$ whose real part is the same as $\psi$ and whose imaginary
part is a solution of first-order partial differential equations
derived from the Cauchy-Riemann condition (see Appendix~\ref{asec_caf}).
Since $2\psi(x,y)=\psi_c(z)+\bar\psi_c(\z)$ and $2\upartial_{\z}\psi=
\upartial_x\psi+\I\upartial_y\psi$ (see Appendix~\ref{asec_ceq}) where
$z=x+\I y$, the resulting lens equation can be written in complex
notation as
\begin{equation}
\label{ple}
\zeta=\zeta_s-\left[\epsilon\f\!(\z)+c\right],
\end{equation}
using a complex analytic function $\epsilon f(z)=\psi_c'(z)-\bar c$,
where $c$ is a complex constant (without the loss of generality,
$\epsilon$ may be taken as real). Note that here and throughout this
paper, the use of primed symbols is exclusively
reserved either for the complex total derivative of an analytic
function or for the ordinary derivative of a real-valued
single-real-variable function, with respect to their argument, while
the argument will be dropped whenever there is little ambiguity.

If $|\epsilon f(z)|\ll1$ for $|z-z_0|\ll1$, where $z_0$ is the image
position of a point mass lens satisfying $\zeta+c=z_0-\z_0^{-1}$, one
can find the image displacement $\delta z=z-z_0$ caused by the small
perturbation of $\epsilon\f(\z)$ for a given source position, by a
series expansion of equation~(\ref{ple}) up to first order of
both $\delta z$ and $\epsilon$,
\begin{equation}
\label{deq}
\delta\zeta=
\upartial_z\zeta_s|_0\delta z+
\upartial_{\z}\zeta_s|_0\delta\z-\epsilon\f\!(\z_0)=
\delta z+\frac{\delta\z}{\z_0^2}-\epsilon\f_0,
\end{equation}
where $\f_0=\f(\z_0)=\overline{f(z_0)}$. If $\J_{s,0}=1-|z_0|^{-4}\ne
0$ (i.e., $|z_0|\ne1$), equation~(\ref{deq}) is invertible for 
$\delta z$. Setting $\delta\zeta=0$, this inversion leads to
\begin{equation}
\label{zdf}
\delta z=
\frac{\epsilon}{\J_{s,0}}\left(\f_0-\frac{f_0}{\z_0^2}\right).
\end{equation}
That is, the new image forms where the perturbation is
counterbalanced by the image of the local linear mapping that
approximates the (unperturbed) lens mapping.

The Jacobian determinant of the lens mapping~(\ref{ple}) is
\begin{equation}
\label{pjd}
\J=1-\frac{1}{|z|^4}+\epsilon
\left(\frac{f^\prime}{\z^2}+\frac{\f^\prime}{z^2}\right)-
\epsilon^2f^\prime\f^\prime=\J_s+
\frac{\epsilon}{|z|^4}\left(z^2f^\prime+\z^2\f^\prime\right)-
\epsilon^2|f^\prime|^2,
\end{equation}
where $\J_s$ is the part of Jacobian determinant that maintains the
same form~(\ref{sjd}) as the point mass lens. The change of the value
of the Jacobian determinant with the perturbation is therefore
understood as the sum of two contributions (up to the linear order);
one due to the direct additional contribution from the perturbation
($\epsilon$-term in eq.~\ref{pjd}) and the other due to the change
of value of $\J_s$ at the new image position
\begin{equation}
\label{dsj}
\delta(\J_s)=\upartial_z\J_s|_0\delta z+
\upartial_{\z}\J_s|_0\delta\z=\frac{2}{|z_0|^4}
\left(\frac{\delta z}{z_0}+\frac{\delta\z}{\z_0}\right).
\end{equation}
When the small-perturbation solution (eq.~\ref{zdf}) is valid, one
may further substitute the solution for $\delta z$;
\begin{equation}
\delta(\J_s)
=\frac{2\epsilon}{\left(1+|z_0|^2\right)|z_0|^2}
\left(\frac{f_0}{\z_0}+\frac{\f_0}{z_0}\right)
=\frac{2\epsilon}{|z_0|^4}\frac{z_0 f_0+\z_0\f_0}{1+|z_0|^2}.
\end{equation}
For this case, the final first order change of the Jacobian
determinant with respect to the point mass lens case is given by the
sum of the two terms
\begin{equation}
\label{dfj}
\delta\J=\J-\J_{s,0}=
\frac{2\epsilon}{|z_0|^4}
\left(\frac{z_0^2f^\prime_0+\z_0^2\f^\prime_0}{2}+
\frac{z_0 f_0+\z_0\f_0}{1+|z_0|^2}\right),
\end{equation}
where $z_0$ is the position of the unperturbed image. In addition, if
$|\delta\J|\ll|\J_{s,0}|=|1-|z_0|^{-4}|$, then the first order change
of the magnification can be found by
\begin{equation}
\label{dfa}
\delta A=-p\frac{\delta\J}{\J_{s,0}^2}=
\frac{2p\epsilon}{2-(|z_0|^4+|z_0|^{-4})}
\left(\frac{z_0^2f^\prime_0+\z_0^2\f^\prime_0}{2}+
\frac{z_0 f_0+\z_0\f_0}{1+|z_0|^2}\right),
\end{equation}
where $p$ is the parity of the image. Provided that the perturbation
does not change the parity of the image, $p=1$ if $|z_0|>1$ and $p=-1$
if $|z_0|<1$. Note that the effect from equation~(\ref{dfa}) is most
likely to be negligible in practice. This is because
$\delta A\sim\epsilon$, and furthermore, for both of the most relevant
regimes of equation~(\ref{dfa}),
$z_0^{-1}(\epsilon^{-1}\delta A)\rightarrow0$ as $z_0\rightarrow0$, and
$z_0(\epsilon^{-1}\delta A)\rightarrow0$ as $z_0^{-1}\rightarrow0$.

\subsection{caustic and critical curve}
\label{ssec_ccc}
As $|z_0|\rightarrow1$ and consequently $\J_{s,0}\rightarrow 0$, the
perturbative solution (eq.~\ref{zdf}) grows and eventually
diverges. That is, the linear perturbation becomes invalid or
incomplete when the image for the point mass lens approaches the Einstein
ring (= the critical curve). This is because the local linear mapping
(eq.~\ref{deq}) that approximates the lens mapping of the point mass
lens becomes projective at the critical point. This implies that the
general perturbative solutions at an arbitrary critical point
require consideration of higher order effects. However, for the limited
case in which the direction of the source displacement coincides with the
projective axis, the solution can be obtained from the linear effect
alone. Moreover, for the point mass lens, the caustic point is
multiply degenerate, and therefore, for source positions near the lens
position (= the caustic point), one can select a valid base point for
the series expansion among any of the critical points on the Einstein
ring that result in the source displacement along the projective axis.

Suppose that the lens equation~(\ref{ple}) is series-expanded at
$z_0=\E^{\I\phi}$. Then, with $\delta z=(\delta r+\I\delta\phi)
\E^{\I\phi}$, the linearized lens equation becomes
\begin{equation}
\label{lle}
\zeta=(\delta r+\I\delta\phi)\E^{\I\phi}+
\frac{(\delta r-\I\delta\phi)\E^{-\I\phi}}{\E^{-2\I\phi}}
-\epsilon\f\!(\E^{-\I\phi})=2\delta r
\E^{\I\phi}-\epsilon\f\!(\E^{-\I\phi}),
\end{equation}
where $\zeta=\delta\zeta$ because $z_0-\z_0^{-1}=0$. Since $\delta r\in
\mathbf R$, the linearized lens equation~(\ref{lle}) has solutions if
$\E^{-\I\phi}[\zeta+\epsilon\f(\E^{-\I\phi})]\in\mathbf R$ or
equivalently $\phi$ is the solution of
\begin{equation}
\label{peq}
F\!(\phi)=\zeta-\E^{2\I\phi}\bar\zeta+\epsilon
\left[\f\!(\E^{-\I\phi})-\E^{2\I\phi}f\!(\E^{\I\phi})\right]=0.
\end{equation}
In general, equation~(\ref{peq}) allows multiple solutions in $[0,
2\upi)$. In principle, for any given expansion base point $z_0=\E^{\I\phi}$,
one can recover multiple solutions of $\delta z$ corresponding to each
of the solutions of equation~(\ref{peq}) if higher order effects
are considered.

From equations~(\ref{pjd}) and (\ref{dsj}), the Jacobian determinant
corresponding to equation~(\ref{lle}) is
\begin{equation}
\label{ljc}
\J=\frac{2}{|z_0|^4}
\left[\frac{\delta z}{z_0}
+\frac{\delta\z}{\z_0}+\frac{\epsilon}{2}
\left(z_0^2f_0^\prime+\z_0^2\f_0^\prime\right)\right]
=4\delta r+\epsilon\left[\E^{2\I\phi}f^\prime\!(\E^{\I\phi})+
\E^{-2\I\phi}\f^\prime\!(\E^{-\I\phi})\right]=
2\E^{-\I\phi}\left[\zeta-\zeta_c(\phi)\right],
\end{equation}
where
\begin{equation}
\label{cau}
\zeta_c(\phi)= 
-\frac{\epsilon}{2}\left[\E^{3\I\phi}f^\prime\!(\E^{\I\phi})+
\E^{-\I\phi}\f^\prime\!(\E^{-\I\phi})+2\f\!(\E^{-\I\phi})\right].
\end{equation}
Here, $\J_{s,0}=1-|z_0|^{-4}=0$ so that $\J=\delta\J$, and also $\J\in
\mathbf R$ if $\phi$ is the solution of equation~(\ref{peq}). Note
that $\J\le1$ for any null-convergent lens system so that
equation~(\ref{ljc}) is only valid if $|\zeta-\zeta_c(\phi)|\ll1/2$.
The total magnification for the given source position can be obtained
by adding the inverse of all Jacobian determinants corresponding to
each solution of equation~(\ref{peq}) in $[0,2\upi)$,
\begin{equation}
\label{mag}
A(\zeta)=
\sum_{\phi_i\in\{0\le\phi<2\upi|F\!(\phi)=0\}}
\frac{1}{2|\zeta-\zeta_c(\phi_i)|}.
\end{equation}

Now note that if there exists $\phi\in[0,2\upi)$ such that
$\zeta_c(\phi)=\zeta$, then $\J(\zeta)=0$ and therefore $A(\zeta)$
diverges. In other words, $\{\zeta_c(\phi)|\phi\in[0,2\upi)\}$ defines
(the linear approximation of) the caustics -- that is, $\zeta_c(\phi)$
is the parametric form of the (linear approximation of the) caustics.
The (parameter for) cusp points formed along the caustics can be found
by solving $d\zeta_c/d\phi=0$ for the parameter $\phi$. Since
\begin{eqnarray}\mbox{\label{dzp}}
\frac{d\zeta_c}{d\phi}&=&-\frac{\epsilon}{2}\left[3\I 
\E^{3\I\phi}f^\prime\!(\E^{\I\phi})+\I
\E^{4\I\phi}f^{\prime\!\prime}\!(\E^{\I\phi})-\I
\E^{-\I\phi}\f^\prime\!(\E^{-\I\phi})-\I
\E^{-2\I\phi}\f^{\prime\!\prime}\!(\E^{-\I\phi})-2\I
\E^{-\I\phi}\f^\prime\!(\E^{-\I\phi})\right]\nonumber\\&=&\frac{\epsilon 
\E^{\I\phi}}{2\I}\left\{\left[3\E^{2\I\phi}f^\prime\!(\E^{\I\phi})
+\E^{3\I\phi}f^{\prime\!\prime}\!(\E^{\I\phi})\right]
-\left[3\E^{-2\I\phi}\f^\prime\!(\E^{-\I\phi})+\E^{-3\I\phi}
\f^{\prime\!\prime}\!(\E^{-\I\phi})\right]\right\}\nonumber\\&=&\epsilon 
\E^{\I\phi}\,\Im\!\left[3\E^{2\I\phi}f^\prime\!(\E^{\I\phi})
+\E^{3\I\phi}f^{\prime\!\prime}\!(\E^{\I\phi})\right]
\end{eqnarray}
where
$\Im[z]:\mathbf C\rightarrow\mathbf R$ is the imaginary part of $z$,
the condition for the parameter $\phi$ to define a cusp point is that
$3\E^{2\I\phi}f'(\E^{\I\phi})+\E^{3\I\phi}f''(\E^{\I\phi})\in\mathbf R$. 

The expression for the corresponding (linear approximation of the)
critical curve can be easily found by setting $\J=0$ in
equation~(\ref{ljc}), that is,
\begin{equation}
\label{lca}
\delta r=\epsilon\chi_1(\phi)=-\frac{\epsilon}{4}
\left[\E^{2\I\phi}f^\prime\!(\E^{\I\phi})+
\E^{-2\I\phi}\f^\prime\!(\E^{-\I\phi})\right].
\end{equation}
Strictly speaking, this only defines the \emph{local} linear
approximation of the critical curve (near $z_0=\E^{\I\phi}$). However,
one may simply consider $(1+\delta r)\E^{\I\phi}$ as a global
expression for the critical curve parametrized by $\phi$,
\begin{equation}
\label{lcr}
z_c(\phi)=\E^{\I\phi}[1+\epsilon\chi_1(\phi)]=\E^{\I\phi}-
\frac{\epsilon}{4}\left[\E^{3\I\phi}f^\prime\!(\E^{\I\phi})+
\E^{-\I\phi}\f^\prime\!(\E^{-\I\phi})\right].
\end{equation}
Then, the image of $z_c(\phi)$ under the linearized lens mapping
\begin{equation}
\label{lcu}
\zeta_c(\phi)=2\E^{\I\phi}\epsilon\chi_1(\phi)-\epsilon\f\!(\E^{-\I\phi}),
\end{equation}
indeed recovers the previous derived expression~(\ref{cau}) for the
caustic.

\section{Inverse symmetry of perturbation potential}
\label{sec_isp}

\subsection{caustic invariant perturbation pair}
If $f(z)$ can be represented by a complex monomial $az^n$ near the
unit circle $|z|=1$ and $|\epsilon a|\ll1$, its caustic can be found
from equation~(\ref{cau}) by
\begin{equation}
\label{mca}
-\frac{2}{\epsilon}\zeta_c(\phi)=
na\E^{3\I\phi}\E^{(n-1)\I\phi}+
n\ab\E^{-\I\phi}\E^{(1-n)\I\phi}+2\ab \E^{-n\I\phi}=
na\E^{(n+2)\I\phi}+(n+2)\ab\E^{-n\I\phi}.
\end{equation}
The resulting expression is a linear combination of two complex
exponentials. Motivated by the relationship between the two exponents
and the scalar coefficients of each exponential, let us consider the
substitution $m=-(n+2)$. Then,
\begin{equation}
na\E^{(n+2)\I\phi}+(n+2)\ab\E^{-n\I\phi}=
(m+2)(-a)\E^{-m\I\phi}+m(-\ab)\E^{(m+2)\I\phi},
\end{equation}
that is, equation~(\ref{mca}) is invariant under the exchange of
$n\leftrightarrow-(n+2)$ accompanied by $a\leftrightarrow-\ab$. In
other words, up to the linear approximation, the shapes of the
caustics due to the perturbations $f(z)=az^n$ and $f(z)=-\ab
z^{-(n+2)}$ are identical. Furthermore, the expression~(\ref{cau}) is
\emph{linear} to terms involving $f$, and therefore, the superposition
principle implies that the caustic resulting from the polynomial
perturbation $f_t(z)=\sum_na_nz^n$ is the same as the one from another
polynomial perturbation $f_p(z)=\sum_n(-\ab_n)z^{-(n+2)}$. Moreover,
one can easily show that equation~(\ref{peq}) is also the same for
both cases, and therefore equation~(\ref{mag}) implies that this
caustic invariance actually extends to the correspondence of the
magnification associated with any source position
in the vicinity of the caustics.

In fact, the pair of `linear caustic invariant perturbations'
(hereafter, LCIP) is also connected by the symmetry between the
corresponding lensing potential. To see this, let us consider two
complex potentials with inverse symmetry; $\psi^t_c(z)$ and
$\psi^p_c(z)=\bar\psi^t_c(z^{-1})$. If $\psi^t_c(z)$ has a convergent
Laurent series expression in a domain containing the neighbourhood of
the unit circle -- since $\psi^t_c(z)$ is an analytic function of $z$
for any null-convergent perturbation, this condition is basically the
restriction to the absence of poles in the neighbourhood of the unit
circle --, then the perturbations due to these potentials are
\begin{equation}
\psi^t_c(z)=\sum_mb_mz^m\ \ \rightarrow\ \
f_t(z)={\psi^t_c}^\prime(z)=\sum_mmb_mz^{m-1}=\sum_{n\ne-1}a_nz^n,
\end{equation}
\begin{equation}
\psi^p_c(z)=\bar\psi^t_c(\frac{1}{z})= 
\sum_m\frac{\bar b_m}{z^m}\ \ \rightarrow\ \
f_p(z)={\psi^p_c}^\prime(z)=\sum_m(-m)\frac{\bar b_m}{z^{m+1}}=
\sum_{n\ne -1}\frac{(-\ab_n)}{z^{n+2}},
\end{equation}
so that they are indeed an LCIP pair. In terms of the real potentials,
the symmetry between the LCIP pair is more straightforward. First, a
general (real) solution of the two-dimensional Laplace equation --
allowing singularities at either the origin or infinity -- in plane polar
coordinates \citep[e.g.,][]{CH62} is, by harmonic expansion,
\begin{eqnarray}\mbox{\label{pty}}
\psi^t&=&a_0\ln r+b_0+\sum_{n=1}^\infty\left[
\left(a_nr^n+\frac{a_{-n}}{r^n}\right)\cos n\phi+
\left(b_nr^n+\frac{b_{-n}}{r^n}\right)\sin n\phi\right]\nonumber\\&=&
\frac{a_0}{2}\ln r^2+b_0+\frac{1}{2}\sum_{n=1}^\infty\left[
\left(a_nr^n+\frac{a_{-n}}{r^n}\right)\left(\E^{n\I\phi}+\E^{-n\I\phi}\right)+
\I\left(b_nr^n+\frac{b_{-n}}{r^n}\right)\left(\E^{-n\I\phi}-\E^{n\I\phi}\right)
\right]\nonumber\\&=&
\frac{a_0}{2}\ln z\z+b_0+\frac{1}{2}\sum_{n=1}^\infty\left[
(a_n-\I b_n)r^n\E^{n\I\phi}+\frac{a_{-n}+\I b_{-n}}{r^n\E^{n\I\phi}}+
(a_n+\I b_n)r^n\E^{-n\I\phi}+\frac{a_{-n}-\I b_{-n}}{r^n\E^{-n\I\phi}}\right],
\end{eqnarray}
where all the constant coefficients, $a_n$'s and $b_n$'s are real. The
corresponding complex potential is easily found as
\begin{equation}
\psi^t_c=a_0\ln z+2b_0+\sum_{n=1}^\infty
\left(\bar c_nz^n+c_{-n}z^{-n}\right),
\end{equation}
and therefore, the complex potential of its LCIP pair is
\begin{equation}
\psi^p_c=a_0\ln z+2b_0+\sum_{n=1}^\infty
\left(c_nz^{-n}+\bar c_{-n}z^n\right),
\end{equation}
where $c_n=a_n+\I b_n$. Finally, the corresponding real potential of
the LCIP pair is
\begin{eqnarray}\mbox{\label{pmp}}
\psi^p=\frac{\psi^p_c+\bar\psi^p_c}{2}&=&
\frac{a_0}{2}\ln z\z+b_0+\frac{1}{2}\sum_{n=1}^\infty\left[
\frac{a_n+\I b_n}{r^n\E^{n\I\phi}}+(a_{-n}-\I b_{-n})r^n\E^{n\I\phi}+
\frac{a_n-\I b_n}{r^n\E^{-n\I\phi}}+(a_{-n}+\I b_{-n})r^n\E^{-n\I\phi}
\right]\nonumber\\&=&
\frac{a_0}{2}\ln r^2+b_0+\frac{1}{2}\sum_{n=1}^\infty\left[
\left(\frac{a_n}{r^n}+a_{-n}r^n\right)\left(\E^{n\I\phi}+\E^{-n\I\phi}\right)+
\I\left(\frac{b_n}{r^n}+b_{-n}r^n\right)\left(\E^{-n\I\phi}-\E^{n\I\phi}\right)
\right]\nonumber\\&=&a_0\ln r+b_0+\sum_{n=1}^\infty\left[
\left(\frac{a_n}{r^n}+a_{-n}r^n\right)\cos n\phi+
\left(\frac{b_n}{r^n}+b_{-n}r^n\right)\sin n\phi\right].
\end{eqnarray} 
In other words, the perturbing parts of the real potentials of the LCIP
pair are related to each other by an inverse symmetry with respect to
the unit circle (i.e., Einstein ring); that is, $\delta\psi^p(r,\phi)=
\delta\psi^t(r^{-1},\phi)$ in plane polar coordinates $(r,\phi)$, or
$\delta\psi^p(\bmath r)=\delta\psi^t(\bmath r/r^2)$ in vector
notation. Harmonic expansion of the potential, in addition, demonstrates
that the presence of an LCIP pair is related to the intrinsic symmetry of
two dimensional harmonic function, that is, the azimuthal structure is
invariant under the radial distance inversion. 

\subsection{caustic invariance and magnification correspondence}
Following the discussion in the previous section, the natural question
arises whether the magnification correspondence between the pair of
potentials with the inverse symmetry extends to sources lying far
from the caustic. For the zeroth order, the answer is affirmative
because the magnification of sources lying far from the caustic is
just small a perturbation on the point mass lens case. However, to
examine whether the correspondence actually extends to the first order
perturbations, one needs to consider the approximations of
equations~(\ref{deq}), (\ref{dfj}), and (\ref{dfa}) that provide one with
the lowest order non-trivial effects due to perturbations for
these cases.

From equation~(\ref{dfa}), the magnification change associated with the
perturbation $f_t(z)=az^n$ is found by
\begin{equation}
\label{mnp}
\delta A_t=
\frac{p\epsilon}{2-(|z_0|^4+|z_0|^{-4})}
\left(n+\frac{2}{1+|z_0|^2}\right)
\left(az_0^{n+1}+\ab\z_0^{n+1}\right).
\end{equation}
On the other hand, the magnification associated with the corresponding
perturbation $f_p(z)=-\ab z^{-(n+2)}$ is
\begin{equation}
\label{mpp}
\delta A_p=
\frac{p\epsilon}{2-(|z_0|^4+|z_0|^{-4})}
\left(n+\frac{2}{1+|z_0|^{-2}}\right)
\left(\frac{a}{\z_0^{n+1}}+\frac{\ab}{z_0^{n+1}}\right).
\end{equation}
Next, one notes that, for a given source position, the symmetry of the
point mass lens implies that the positive and negative parity images
should be related by $z_0\leftrightarrow-\z_0^{-1}$. Hence, the linear
changes in magnification on alternative parity images by a pair of
perturbations $f_t(z)$ and $f_p(z)$ are symmetric for even $n$ -- an
odd potential -- while they are antisymmetric for odd $n$ -- an even
potential. In other words, the pair of perturbations $f_t(z)=
az^n$ and $f_p(z)=(-1)^{n+1}\ab z^{-(n+2)}$ yields correspondent
linearly perturbed magnifications for the source position far from the
caustics. Using the same argument in the previous section, one may
generally conclude that this pair is in fact related, in terms of
potentials, with a property that $\psi^p(\bmath r)=
-\psi^t(-\bmath r/r^2)$ or $\psi^p_c(z)=-\bar\psi^t_c(-z^{-1})$. In
addition, one may also state that the magnification correspondence
between the LCIP pair of odd potentials is stronger than that of even
potentials.

\section{Examples}
\label{sec_egs}

\subsection{planetary perturbations}
\label{ppp}
\citet{Bo99} showed the perturbative analysis provides an ideal
method to study microlensing from planetary perturbations. Here,
some well-known degeneracies of planetary microlensing are
reexamined using the perturbative approach. In general,
microlensing by a star with a planet can be described by the lens
equation,
\begin{equation}
\label{pln}
\zeta=z-\frac{1}{\z}-\frac{q}{\z-\z_\mathrm{p}}.
\end{equation}
Here, the position of star is chosen as the coordinate origin, and
Einstein ring radius corresponding to the stellar mass alone as the
unit of angular measurements. The projected angular location of the
planet is given by $z_\mathrm{p}$, while $q$ is the mass ratio of the
planet to the star. From the comparison to equation~(\ref{ple}), if
$0<q\ll1$ (and $|z-z_\mathrm{p}|\gg q$), equation~(\ref{pln}) can
be considered as a point mass lens under a small null-convergent
perturbation that comes from
\begin{equation}
f(z)=(z-z_\mathrm{p})^{-1}\,;\ \ \ f^\prime(z)=-(z-z_\mathrm{p})^{-2},
\end{equation}
which is analytic everywhere except $z=z_\mathrm{p}$ where it forms a
pole. Then, if $z_\mathrm{p}$ is sufficiently far from the circle
$|z|=1$, equation~(\ref{cau}) provides a parametric representation for
the linear approximation of the (central) caustics;
\begin{equation}
\label{pca}
\frac{2}{q}\zeta_c=\frac{\E^{3\I\phi}}{(\E^{\I\phi}-z_\mathrm{p})^2}+
\frac{\E^{-\I\phi}}{(\E^{-\I\phi}-\z_\mathrm{p})^2}-
\frac{2}{\E^{-\I\phi}-\z_\mathrm{p}},
\end{equation}
where $\phi\in[0,2\upi)$ is a parameter. To examine the degeneracies
of planetary microlensing related to these caustics, it is helpful
to further manipulate the last two terms in the equation~(\ref{pca})
algebraically, that is,
\begin{equation}
\label{man}
\frac{\E^{-\I\phi}}{(\E^{-\I\phi}-\z_\mathrm{p})^2}
-\frac{2}{\E^{-\I\phi}-\z_\mathrm{p}}=
\frac{\z_\mathrm{p}^{-2}\E^{\I\phi}}{(\E^{\I\phi}-\z_\mathrm{p}^{-1})^2}+
\frac{2\z_\mathrm{p}^{-1}\E^{\I\phi}}{\E^{\I\phi}-\z_\mathrm{p}^{-1}}=
\E^{\I\phi}\left(\frac{\z_\mathrm{p}^{-1}}{\E^{\I\phi}-
\z_\mathrm{p}^{-1}}+1\right)^2-\E^{\I\phi}=
\frac{\E^{3\I\phi}}{(\E^{\I\phi}-\z_\mathrm{p}^{-1})^2}-\E^{\I\phi}.
\end{equation}
Then, combining equations~(\ref{pca}) and (\ref{man}) leads to a more
symmetric representation of the linear approximation of the caustics;
\begin{equation}
\label{pcu}
\frac{2}{q}\zeta_c=\frac{\E^{3\I\phi}}{(\E^{\I\phi}-z_\mathrm{p})^2}+
\frac{\E^{3\I\phi}}{(\E^{\I\phi}-\z_\mathrm{p}^{-1})^2}-\E^{\I\phi}=
\E^{\I\phi}\left[
\frac{1}{(1-z_\mathrm{p}\E^{-\I\phi})^2}+
\frac{1}{(1-\z_\mathrm{p}^{-1}\E^{-\I\phi})^2}-1\right].
\end{equation}
Therefore, up to linear order, the shapes of the caustics due
to the perturbation from the planet lying at $z_\mathrm{p}$ and the
one at $\z_\mathrm{p}^{-1}$ are identical. Note that $z_\mathrm{p}$
and $\z_\mathrm{p}^{-1}$ both have the same argument but that their
norms are related inversely to each other. It is also straightforward
to show that this symmetry further implies the magnification
degeneracy between these two planet positions through
equation~(\ref{mag}). Compare this to the degeneracy identified by
\citet{GG97} as their second class of discrete degeneracies of
planetary microlensing -- ``{\it whether the planet lies closer to or
farther from the star than does the position of the image that it is
perturbing}.'' Note that this degeneracy also extends to systems of
any number of multiple planets since the superposition
principle applies to the linear approximation.

If one compares the actual central caustics of a planetary
microlens to its linear approximation in equation~(\ref{pcu}), one
finds that deviations between them take place for smaller $q$ with
$|z_\mathrm{p}|<1$ than $|z_\mathrm{p}|>1$, and therefore that the
$z_\mathrm{p}$-$\z_\mathrm{p}^{-1}$ degeneracy appears not to be
strict for $q\ga10^{-2}$. This is because the magnitude of the actual
perturbation term ($|qf|\sim q$) for $|z_\mathrm{p}|<1$ grows faster
than the one for $|z_\mathrm{p}|>1$, which grows $|qf|\sim
q|z_\mathrm{p}|^{-1}$ for $|z_0|\sim1$. To find a better
approximation of the caustics for close-in ($|z_\mathrm{p}|<1$)
planetary systems, one may need to include the second order effects of
the perturbation (see Appendix~\ref{asec_spc}). However, if $0<
|z_\mathrm{p}|\ll1$, it is possible to find an alternative
perturbative approach for which the leading perturbation term behaves
as $\sim q|z_\mathrm{p}|^2$, and therefore the linear approximation
that can be used for larger $q$. Similarly, one can devise a different
description of the system in which the leading perturbation term
behaves as $\sim q|z_\mathrm{p}|^{-2}$, which can be applied for
wide-separation ($|z_\mathrm{p}|\gg1$) planetary systems. Furthermore,
one can also show that these two types of perturbative description of
system, in fact, lead to a certain pair of LCIP, which may be
understood as a natural extension of the
$z_\mathrm{p}$-$\z_\mathrm{p}^{-1}$ degeneracy.

\subsection{extreme binary lens}
\label{ebp}
It was \citet{Do99b} who noticed that the approximation of binary
lenses with extreme separations by certain classes of perturbed point
mass lens systems reveals the underlying connection between
various binary lens systems. \citet{Al02} tried explicit but rather
limited calculations based on the perturbative approach to demonstrate
the presence of the magnification degeneracy between two types of
binary lens systems. Here, this magnification degeneracy is studied in
a more general way as an archetypal example of the LCIP pair.

Let us first think of the situation in which the lens equation is
given by equation~(\ref{mle}) with two component masses. By
series-expanding the deflection term caused by one of the masses at the
location of the other mass;
\begin{equation}
\zeta=z-\frac{q_1}{\z-\lb_{z_1}}-\frac{q_2}{\z-\lb_{z_2}}=
z-\frac{q_1}{\z-\lb_{z_1}}+\frac{q_2}{\lb_{z_2}-\lb_{z_1}}
\sum_{k=0}^\infty\frac{(\z-\lb_{z_1})^k}{(\lb_{z_2}-\lb_{z_1})^k}
=z-\frac{q_1}{\z-\lb_{z_1}}+\frac{q_2}{\lb_{z_2}-\lb_{z_1}}+
\sum_{k=1}^\infty\frac{q_2(\z-\lb_{z_1})^k}{(\lb_{z_2}-\lb_{z_1})^{k+1}},
\label{wbe}
\end{equation}
where $l_{z_1}$ and $l_{z_2}$ are the positions of the two mass points. 
The radius of the convergence for the infinite series in
equation~(\ref{wbe}) is given by $|z-l_{z_1}|<|l_{z_2}-l_{z_1}|$.
After some substitutions of symbols,
\begin{equation}
w=q_1^{-1/2}(z-l_{z_1})\,;\ \ \
\omega=\frac{1}{\sqrt{q_1}}
\left(\zeta-l_{z_1}-\frac{q_2}{\lb_{z_2}-\lb_{z_1}}\right)
=q_1^{-1/2}(\zeta-l_{z_1})-\frac{q_2}{q_1}(\lb_{w_2}-\lb_{w_1})^{-1},
\end{equation}
which is basically the choice of new coordinate origins [$l_{z_1}$
for the image space and $l_{z_1}-q_2(\lb_{z_1}-\lb_{z_2})^{-1}$ for
the source space] and Einstein ring radius of the mass at
$l_{z_1}$ as the new unit of angular measurements, it is easy to see
that lens equation~(\ref{wbe}) is a specific example of lens
equation~(\ref{ple}) for a point mass lens under perturbations,
\begin{equation}
\label{crl}
\omega=w-\frac{1}{\w}+\sum_{k=1}^\infty\gamma_{k-1}\w^k,
\end{equation}
where
\begin{equation}
\gamma_k=\frac{q_2q_1^{k/2}}{(\lb_{z_2}-\lb_{z_1})^{k+2}}=
\frac{q_2}{q_1}(\lb_{w_2}-\lb_{w_1})^{-(k+2)}.
\end{equation}
Here, $l_{w_k}=q_1^{-1/2}(l_{z_k}-l_{z_1})$, that is, the lens
location in this new coordinate system. In particular,
\begin{equation}
\gamma_0=\frac{q_2}{q_1}(\lb_{w_2}-\lb_{w_1})^{-2}\,;\ \ \
\gamma_k=\frac{\gamma_0}{(\lb_{w_2}-\lb_{w_1})^k}.
\end{equation}
If $|\gamma_0|\ll1$, then, the circle $|w|=1$ lies well within the
region of the convergence of the infinite series, and also the
perturbative approach for the caustics in the previous section is
valid. If one considers the linear approximation alone, the
perturbation series may be truncated after the second term. Note that,
here, the second order effect of the leading term in the series
[$|\gamma_0^2|\sim|l_{w_2}-l_{w_1}|^{-4}$] is higher order than
the linear effect from the second term
[$|\gamma_1|\sim|l_{w_2}-l_{w_1}|^{-3}$]. In effect, the lens
equation~(\ref{crl}) describes the system as a point mass lens lying
in the gravitational tidal field due to masses that are well
outside of the region around its Einstein ring.

Next, let us consider the multipole expansion of the deflection terms
of the lens equation~(\ref{mle}) at some point $z_\mathrm{c}$;
\begin{eqnarray}\mbox{\label{cbe}}
\lefteqn{
\zeta=z-\frac{q_1}{\z-\lb_{z_1}}-\frac{q_2}{\z-\lb_{z_2}}=
z-\frac{q_1}{\z-\z_\mathrm{c}}
\sum_{k=0}^\infty\frac{(\lb_{z_1}-\z_\mathrm{c})^k}{(\z-\z_\mathrm{c})^k}-
\frac{q_2}{\z-\z_\mathrm{c}}
\sum_{k=0}^\infty\frac{(\lb_{z_2}-\z_\mathrm{c})^k}{(\z-\z_\mathrm{c})^k}
}\nonumber\\&&
=z-\frac{q_1+q_2}{\z-\z_\mathrm{c}}-
\frac{q_1\lb_{z_1}+q_2\lb_{z_2}-(q_1+q_2)\z_\mathrm{c}}{(\z-\z_\mathrm{c})^2}-
\sum_{k=1}^\infty
\frac{q_1(\lb_{z_1}-\z_\mathrm{c})^{k+1}+q_2(\lb_{z_2}-\z_\mathrm{c})^{k+1}}{(\z-\z_\mathrm{c})^{k+2}}.
\end{eqnarray}
Here, the radius of the convergence for the infinite series is given
by $|z-z_\mathrm{c}|>\max(|l_{z_1}-z_\mathrm{c}|,|l_{z_2}-z_\mathrm{c}|)$.
If one chooses $z_\mathrm{c}$
to be the centre of mass of the system,
\begin{equation}
z_\mathrm{c}=\frac{q_1}{q_1+q_2}l_{z_1}+\frac{q_2}{q_1+q_2}l_{z_2}=
\frac{q_1l_{z_1}+q_2l_{z_2}}{q_1+q_2},
\end{equation}
then the dipole term vanishes. Subsequently, substituting symbols
via
\begin{equation}
w=(q_1+q_2)^{-1/2}(z-z_\mathrm{c})\,;\ \ \
\omega=(q_1+q_2)^{-1/2}(\zeta-z_\mathrm{c}),
\end{equation}
which is the choice of the centre of the mass $z_\mathrm{c}$ being the
coordinate origin and Einstein ring radius corresponding to the total
mass of the system being the unit of angular measurements, one finds
equation~(\ref{cbe}) to be consistent with being another example
of a point mass lens under perturbations (eq.~\ref{ple}),
\begin{equation}
\label{qpl}
\omega=w-\frac{1}{\w}-\sum_{k=1}^\infty\hat Q_{2^{k+1}}\w^{-(k+2)},
\end{equation}
where
\begin{eqnarray}
\hat Q_{2^k}&=&
\frac{q_1(\lb_{z_1}-\z_\mathrm{c})^k+q_2(\lb_{z_2}-\z_\mathrm{c})^k}{(q_1+q_2)^{(k/2)+1}}
=\frac{q_1l_{w_1}^k+q_2l_{w_2}^k}{q_1+q_2}\\&=&
\frac{q_1q_2^k(\lb_{z_1}-\lb_{z_2})^k+q_2q_1^k(\lb_{z_2}-\lb_{z_1})^k}
{(q_1+q_2)^{k/2}(q_1+q_2)^{k+1}}=
\frac{q_1(-q_2)^k+q_1^kq_2}{(q_1+q_2)^{k+1}}(\lb_{w_2}-\lb_{w_1})^k
=\frac{-\left[\sum_{j=1}^{k-1}q_1^j(-q_2)^{k-j}\right]}{(q_1+q_2)^k}
(\lb_{w_2}-\lb_{w_1})^k
\end{eqnarray}
are basically multipole moments. Here, again $l_{w_k}=(q_1+q_2)^{-1/2}
(l_{z_k}-z_\mathrm{c})$ is the lens location in the new coordinate system, but
the relation to the original representation differs from the previous
case because of the difference of the new coordinate system.
The coefficient for the leading term (the
quadrupole moment) and the following higher order moments can also be
written as
\begin{equation}
\hat Q_4=\frac{q_1q_2}{(q_1+q_2)^2}(\lb_{w_2}-\lb_{w_1})^2\,;\ \ \
\hat Q_{4\cdot2^k}=\hat Q_4\frac{\sum_{j=0}^kq_1^j(-q_2)^{k-j}}
{(q_1+q_2)^k}(\lb_{w_2}-\lb_{w_1})^k=\hat Q_4
\frac{q_1^{k+1}-(-q_2)^{k+1}}{(q_1+q_2)^{k+1}}(\lb_{w_2}-\lb_{w_1})^k.
\end{equation}
Again, if $|\hat Q_4|\ll1$, one finds not only that the infinite
series converges at $|w|=1$ but also that the perturbation series is small
enough for the linear approximation of the caustics given in
equation~(\ref{cau}) to be reasonably valid. Similar to the previous
case, one finds that the second order effect of the leading term in
the series [$|\hat Q_4|^2\sim|l_{w_2}-l_{w_1}|^4$] is a higher
order effect than the linear effect of the second term [$|\hat Q_8|
\sim|l_{w_2}-l_{w_1}|^3$] and so one may truncate the
perturbation series after the second term when considering of the
linear approximation.

Finally, if one compares equations~(\ref{crl}) and (\ref{qpl}),
one immediately discovers that the two become a strict LCIP pair if
$\bar\gamma_{k-1}=\hat Q_{2^{k+1}}$ ($k=1,2,\ldots$). However, as
noted above for each case, one may truncate each series after its
second term when one is considering only linear effects. Hence, the
conditions that $\bar\gamma_0=\hat Q_4$ and $\bar\gamma_1=\hat Q_8$,
in fact, suffice to define an LCIP pair from equations~(\ref{crl})
and (\ref{qpl}). Following the usual convention of parameter definition
for the binary lens, that is, $\ds=(q_1+q_2)^{-1/2}(l_{z_2}-l_{z_1})$
as the projected separation between two component in units of the Einstein
ring corresponding to the total mass of the system and $\qr=q_2/q_1>0$ as
the mass ratio between the two components, one can rewrite the leading
coefficients of the series
\begin{equation}
\label{tdl}
\bar\gamma_0=\frac{\qr_t}{(1+\qr_t)\ds_t^2}\,;\ \ \
\bar\gamma_1=\frac{\bar\gamma_0}{(1+\qr_t)^{1/2}\ds_t}
\end{equation}
\begin{equation}
\label{mpl}
\hat Q_4=\frac{\qr_p\bar\ds_p^2}{(1+\qr_p)^2}\,;\ \ \
\hat Q_8=\hat Q_4\frac{(1-\qr_p)\bar\ds_p}{1+\qr_p},
\end{equation}
where the subscripts `$_t$' and `$_p$' are used to distinguish the
parameters associated with the tidal approximation (for $|\ds_t|\gg
1$) and the multipole expansion (for $|\ds_p|\ll 1$). Then, the
condition for two system to be an LCIP pair becomes
\begin{equation}
(1+\qr_t)(\ds_t\bar\ds_p)^2=\qr_t\frac{(1+\qr_p)^2}{\qr_p}\,;\ \ \
(1+\qr_t)^{1/2}\ds_t\bar\ds_p=\frac{1+\qr_p}{1-\qr_p}.
\end{equation}
This condition can be used to find the system ($\qr_t$, $\ds_t$)
that is degenerate in the first order to the system ($\qr_p$, $\ds_p$);
\begin{equation}
\label{one}
\qr_t=\frac{\qr_p}{(1-\qr_p)^2}\,;\ \ \
\ds_t=\frac{1}{\bar\ds_p}\frac{1+\qr_p}{(1-\qr_p+\qr_p^2)^{1/2}},
\end{equation}
or the system ($\qr_p$, $\ds_p$) to the system ($\qr_t$, $\ds_t$);
\begin{equation}
\label{two}
\qr_p=\frac{1+2\qr_t-(1+4\qr_t)^{1/2}}{2\qr_t}
=1-\frac{(1+4\qr_t)^{1/2}-1}{2\qr_t}\,;\ \ \
\ds_p=\frac{1}{\bar\ds_t}\left(\frac{1+4\qr_t}{1+\qr_t}\right)^{1/2}.
\end{equation}
Algebraically, there is a second solution for the inversion of the
equation~(\ref{one}). However, the second solution is related to
equation~(\ref{two}) by $\qr_p\leftrightarrow\qr_p^{-1}$ and
$\ds_p\leftrightarrow-\ds_p$, which results in the identical system
because each relationship on its own corresponds to a 180\degr\ rotation
of the system (with respect to the centre of mass) and so the combination
of the two becomes the identity transformation. Note that, if $\qr\ll 1$,
then $\qr_t\approx\qr_p+2\qr_p^2$, $\qr_p\approx\qr_t-2\qr_t^2$, and
$\ds_t\bar\ds_p\approx 1+(3/2)\qr$. That is, this LCIP pair is reduced to
the $z_\mathrm{p}$-$\z_\mathrm{p}^{-1}$ degeneracy of planetary perturbations.

\subsubsection{central caustic}

\begin{figure}
\plot{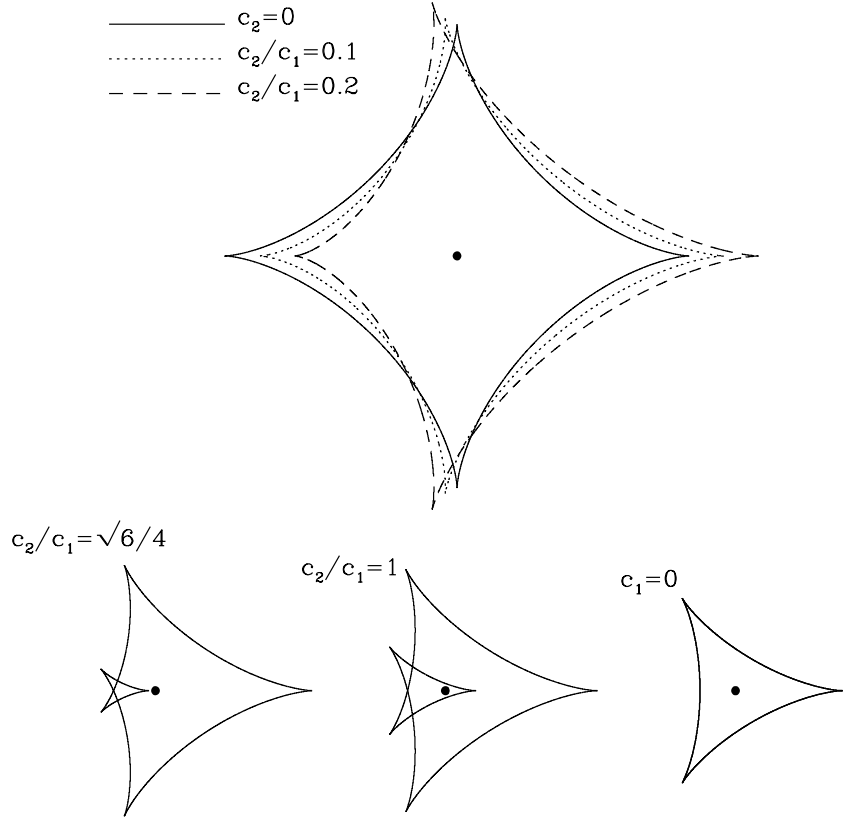}
\caption{
Linear approximation for the caustics of quadr-octu-pole lens given by
eq.~(\ref{ccc}). The real axis (the horizontal direction) is chosen
such that both coefficients are real (i.e., $\phi_1=\phi_2=0$). The
dots represent the coordinate centre. In the top panel, the shape
change of the tetracuspoid caustics with varying $s=c_2/c_1$ at fixed $c_1$ is
represented. The bottom panels, on the other hand, show hexacuspoid (or
tricuspoid if $c_1=0$) caustics with fixed values of eq.~(\ref{are}) -- it
is \emph{not} the same as an area since the curves cross themselves.
}\label{oql}
\end{figure}

Detailed discussion regarding the analysis of the shape of the
(central) caustics of extreme binary lens is available in the
literature \citep[e.g.,][]{Do99b,Bo00}. Here, the basic properties of
these caustics are reexamined as an example of the simplest LCIP
caustics, essentially controlled by two parameters.

Suppose that a point mass lens is perturbed by a null-convergent
perturbation of the form of $f=-c_1z-c_2z^2$. From
equation~(\ref{mca}), the linear approximation of the resulting
caustic is
\begin{equation}
\label{ccc}
\zeta_c=\frac{1}{2}\left(
c_1\E^{3\I\phi}+3\bar c_1 \E^{-\I\phi}+
2c_2\E^{4\I\phi}+4\bar c_2 \E^{-2\I\phi}\right)=
\frac{|c_1|}{2}\E^{\I\phi_1}\left[
\E^{3\I(\phi-\phi_1)}+3\E^{-\I(\phi-\phi_1)}\right]+
|c_2|\E^{\I\phi_2}\left[
\E^{4\I(\phi-\phi_2)}+2\E^{-2\I(\phi-\phi_2)}\right],
\end{equation}
where $-2\phi_1$ and $-3\phi_2$ are the arguments of $c_1$ and
$c_2$ respectively (i.e., $c_1=|c_1|\E^{-2\I\phi_1}$ and $c_2=|c_2|
\E^{-3\I\phi_3}$). Up to the same order, this caustic is identical to
that caused by its LCIP, $f=\bar c_1z^{-3}+\bar c_2z^{-4}$. In
general, $\E^{\I(\phi_1-\phi_2)}$ is not necessarily real. However, if
one restricts to the approximation for extreme binary lenses, one finds
that $\phi_1=\phi_2$. That is, $c_1=\bar\gamma_0$ and $c_2/c_1=
\bar\gamma_1/\bar\gamma_0$ for the tidal approximation, or $c_1=
\hat Q_4$ and $c_2/c_1=\hat Q_8/\hat Q_4$, and therefore, from
equations~(\ref{tdl}) and (\ref{mpl}), both $\phi_1$ and $\phi_2$ are
the argument of $\ds$ (i.e., $\ds/|\ds|=\E^{\I\phi_1}=\E^{\I\phi_2}$).
Then,
\begin{equation}
\zeta_c \E^{-\I\phi_c}=
2|c_1|\left[\cos^3(\phi-\phi_c)-\I\sin^3(\phi-\phi_c)\right]+
|c_2|\left\{4\cos^2(\phi-\phi_c)\cos[2(\phi-\phi_c)]-1
-4\I\sin^2(\phi-\phi_c)\sin[2(\phi-\phi_c)]\right\},
\label{cri}
\end{equation}
where $\phi_c=\phi_1=\phi_2$. If $c_2=0$ (and the real axis is
chosen such that $\phi_c=0$), the expression is reduced to the one
derived by \citet{Ko87b}. To find the cusp points (eq.~\ref{dzp}),
\begin{equation}
\label{cpp}
\frac{d\zeta_c}{d\phi}=\frac{\E^{\I\phi}}{2\I}\left[
3\left(\bar c_1 \E^{-2\I\phi}-c_1\E^{2\I\phi}\right)+
8\left(\bar c_2 \E^{-3\I\phi}-c_2\E^{3\I\phi}\right)\right]=
\E^{\I\phi}\left\{
3|c_1|\sin[2(\phi-\phi_1)]+8|c_2|\sin[3(\phi-\phi_2)]\right\}=0.
\end{equation}
Here, if one still assumes that $\phi_c=\phi_1=\phi_2$, one
immediately finds that a cusp forms with $\phi-\phi_c=n\upi$, where $n$
is integer. For these cusps, $\Re[\zeta_c \E^{-\I\phi_c}]=(-1)^n2|c_1|
+3|c_2|$ and $\Im[\zeta_c \E^{-\I\phi_c}]=0$ so that they lie along the
axis with the angle of $\phi_c$ and are separated by $4|c_1|$.
Here, similar to $\Im[z]$, the imaginary part of $z$,
$\Re[z]:\mathbf C\rightarrow\mathbf R$ is the real part of $z$ (i.e.,
$\Re[z]=(z+\z)/2$, $\Im[z]=\I(\z-z)/2$, $z=\Re[z]+\I\Im[z]$).
In addition, one can show that there is either one more solution of
equation~(\ref{cpp}) in $0<\phi-\phi_c<\upi$ if $4|c_2|<|c_1|$ or two
solutions in the same interval if $4|c_2|>|c_1|$, and therefore that
the caustic has four ($4|c_2|\le|c_1|$)\footnote{Strictly speaking,
at $4|c_2|=|c_1|$,
the caustics develop a butterfly catastrophe, and therefore, it has
three simple cusps plus one butterfly catastrophe.}
or six cusps ($4|c_2|>|c_1| >0$)
unless $c_1=0$. If $c_1=0$ and $c_2\ne0$, it becomes tricuspoid --
three cusps -- since the $\phi$-period for eq.~(\ref{ccc}) then
becomes $\upi$ rather than $2\upi$. However, hexacuspoid caustics generated
by equation~(\ref{ccc}) with $4|c_2|>|c_1|$ are doubly-wound
self-intersecting curves (Fig.~\ref{oql}), which cannot be the caustics
of any real pure binary lens (note that certain ternary lenses may be
approximated by an octupole dominated perturbation series and thus
consistent with hexacuspoid caustics although $\phi_1$ and $\phi_2$ may
not be the same any more). Hence, the tetracuspoid condition may be regarded as
one of the limit for the approximation of an extreme binary lens by a
perturbation series. The area enclosed by a tetracuspoid caustic
can be found by\footnote{Note that the area two-form can be written as
$dx\wedge dy=(d\z\wedge dz)/(2\I)=d(\z dz)/(2\I)$. Hence, from the
fundamental theorem of multivariative calculus, the area bounded by $\{z\}$
can be found to be $(2\I)^{-1}\int d(\z dz)=(2\I)^{-1}\oint\z dz$.}
\begin{equation}
\label{are}
S=\frac{1}{2\I}\oint\bar\zeta_c\D\zeta_c
=\frac{1}{2\I}\int_0^{2\upi}\bar\zeta_c\frac{d\zeta_c}{d\phi}\D\phi
=-\left(\frac{3}{2}\upi|c_1|^2+4\upi|c_2|^2\right),
\end{equation}
where the negative sign is caused by the orientation of the current
$\phi$-parametrization of the caustic. With the restriction that
$|c_2/c_1|\ll1/4$, one may regard $c_1$ as the main parameter
controlling the size of the caustics while $s=c_2/c_1$ is an asymmetry
shape parameter for the caustics. In Fig.~\ref{oql}, the change of
the caustics shape with varying value of $s$ at fixed $c_1$ is
represented. The dimension is intentionally omitted since it can be
linearly scaled with $c_1$. In particular, the distance between two
cusps on the horizontal direction is given by $4|c_1|$ and is independent
of $s$. For completeness, examples of the caustics with $s>0.25$ are
also shown in the bottom of Fig.~\ref{oql} although they are not
applicable for the approximation of the caustics of any binary lens.

\begin{figure}
\plot{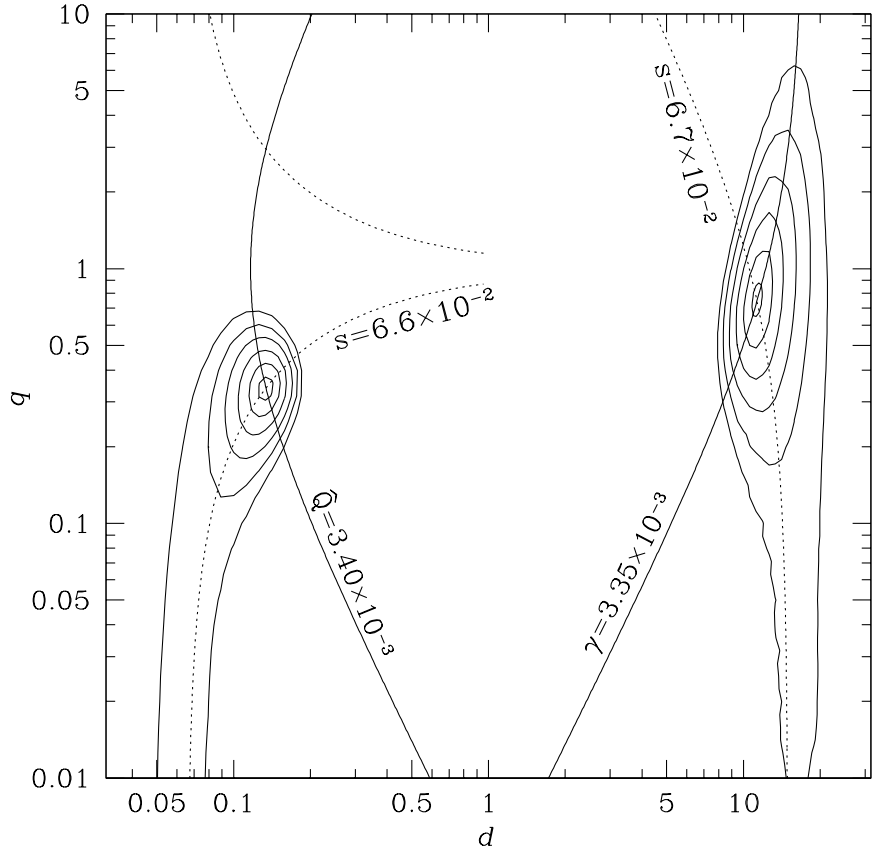}
\caption{
Fig.~2 of \citet{Al02} along with additional dotted lines showing
the curves of constant shape parameter $s$ with the value for their
best-fit models. For the extreme close binary, it corresponds to
$|s|=|\hat Q_8/\hat Q_4|=(1-\qr_p)(1+\qr_p)^{-1}|\ds_p|$ while it is
given by $|s|=|\gamma_1/\gamma_0|=(1+\qr_t)^{-1/2}|\ds_t|^{-1}$ for
the extreme wide binary. 
(The original version \copyright\ 2002 the American Astronomical Society).
}\label{alc}
\end{figure}

The fact that the shape of the central caustics of extreme binary lenses
is controlled not by the leading coefficient of the perturbation
series but by the ratio of its two first coefficients also implies
that, when the wing of the lightcurve is not well observed, there may
exist a continuous degeneracy of the binary lens model running along
the (almost) constant shape parameter. While the behaviour of the
highly-magnified part of lightcurve when the source is near the
central caustic can constrain the structure of the caustic well
enough to determine the shape parameter, the constraints on the relative
size of the caustics with respect to the Einstein ring require the precise
determination of the overall time-scale. Since the magnification near
the centre is essentially scale-free ($A\sim|\zeta|^{-1}$; c.f.,
eq.~\ref{mag}) when the size of the caustic is sufficiently small compared
to the distance between the caustic and the source, one may rescale
the whole lightcurve by changing the blend fractions. Therefore,
without the detailed constraints from the wing, the overall scale
factor is left to be unconstrained and the model exhibits a
continuous degeneracy with strongly correlated time-scale, peak
magnification, blending, and the size parameter of the caustics (which
is the leading coefficient of the perturbation series). One can find an
archetypal demonstration of this degeneracy in the modelling of the
lightcurve of MACHO 99-BLG-47 \citep{Al02}. In Fig.~\ref{alc}, one
can not only find an example of a two-fold LCIP degeneracy but also the
continuous degeneracy with the constant shape parameter.

It is notable that, as $\qr\rightarrow0$, the shape parameter is
mainly controlled by $\ds$ alone (i.e., $s\sim\bar\ds_p$ or $s\sim
\ds_t^{-1}$). In fact, if $\qr\ll1$, the approximation of the
caustic by equation~(\ref{pcu}) can be valid, and one can immediately
find that the shape of the caustic is completely (up to linear order)
determined by the planetary position $z_\mathrm{p}$ while the planetary mass
ratio $q$ becomes an overall linear scale. Moreover, Taylor-series
expansion of equation~(\ref{pcu}) either for $|z_\mathrm{p}|\ll1$:
\begin{equation}
\zeta_c(\phi)=qz_\mathrm{p}+\frac{q}{2}\left[
3z_\mathrm{p}^2\E^{-\I\phi}+\z_\mathrm{p}^2\E^{3\I\phi}+
4z_\mathrm{p}^3\E^{-2\I\phi}+2\z_\mathrm{p}^3\E^{4\I\phi}+\mathcal O
(|z_\mathrm{p}|^4)\right],
\end{equation}
or for $|z_\mathrm{p}|\gg1$:
\begin{equation}
\zeta_c(\phi)=\frac{q}{\z_\mathrm{p}}+\frac{q}{2}\left[
\frac{\E^{3\I\phi}}{z_\mathrm{p}^2}+3\frac{\E^{-\I\phi}}{\z_\mathrm{p}^2}+
2\frac{\E^{4\I\phi}}{z_\mathrm{p}^3}+4\frac{\E^{-2\I\phi}}{\z_\mathrm{p}^3}+
\mathcal O(|z_\mathrm{p}|^{-4})\right]
\end{equation}
approaches equation~(\ref{ccc}) with the shape parameter given by
$s=\z_\mathrm{p}$ (for $|z_\mathrm{p}|\ll1$) or $s=z_\mathrm{p}^{-1}$ (for $|z_\mathrm{p}|\gg1$) as expected.

\subsubsection{planetary caustic}
Note that the condition for the series expansion~(\ref{crl}) to be
valid for $|w|\sim1$ is that $|l_{w_2}-l_{w_1}|=(1+\qr)^{1/2}|\ds|\gg
1$. That is, if $\qr\gg1$, then the tidal approximation can be used
even if $|\ds|\sim1$. However, for this case, one can easily notice
that $|\gamma_0|=(1+\qr^{-1})^{-1}|\ds|^{-2}\sim1$ (eq.~\ref{tdl})
so that the lensing behaviour of these systems cannot be properly
described by the perturbative analysis although the subsequent higher
order terms may be ignored (or viewed as perturbations on a lens
with linear shear). If one truncates the series after the leading term,
one finds that the system modelled by the lens equation~(\ref{crl}) is
a point mass lens subject to a constant external shear given by
$-\gamma_0$,\footnote{\footnotesize
The negative sign follows from the definition
given in Appendix~\ref{asec_ceq}, which conforms to the usual convention
found in weak lensing literature -- the direction of the
shear to be the direction of the eigenvector associated with the positive
eigenvalue of $\bmath\gammaup$. Assuming $\ds\in\mathbf R$,
`$-\gamma_0$' is negative real (eq.~\ref{crl}) \footnotesize
so that, according to
this definition, the direction of the external shear due to the
companion at the the position of the planetary caustics is along the
imaginary axis, that is, the shear runs perpendicular to the axis
joining two masses.} 
which is generally referred to as a Chang-Refsdal lens
after \citet{CR79,CR84}. While the lensing behaviour of Chang-Refsdal
lenses is for the most part analytically tractable, its study is beyond the
scope of the current monograph. Here, it is simply noted that, for
planetary microlensing, provided that the planet lies sufficiently far
from the Einstein ring, there exist two types of caustics; one that can be
approximated by equation~(\ref{pcu}) -- the central caustic, and the
other, which is basically the caustic of Chang-Refsdal lens perturbed
by the higher order tidal-effect terms -- the planetary caustic. In
general, the lensing effects due to planetary caustics are rather
smaller than those associated with the central caustics so that they
usually appear as small additional signals on the main lensing event
\citep{GL92,GG97}.

\section{Discussion}
\label{sec_dis}

\subsection{What about the critical curves?}
\label{ssec_wcc}

\begin{figure}
\plot{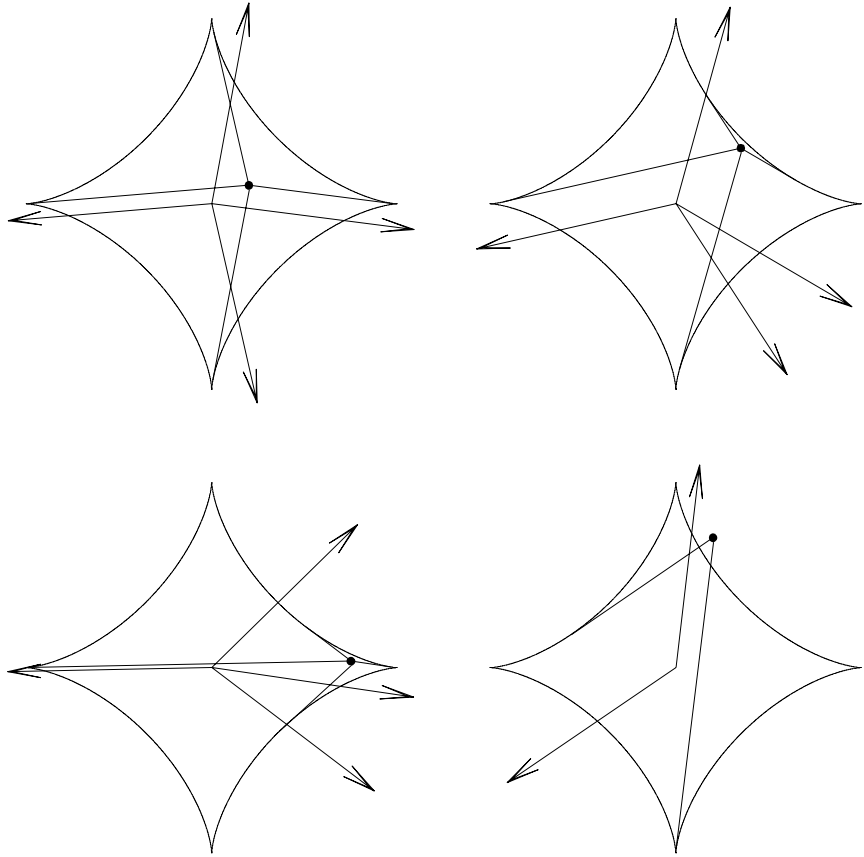}
\caption{
Geometrical determination of the azimuthal positions and magnifications of
images with a linear approximation of the caustics. For this example,
the two cusps lying along the horizontal direction are positive so
that the tangent to a caustic point on the bottom half points upwards
and vice versa. For a given shape the caustics, if one can draw a
tangent to the caustics through a given source position (represented
by a dot), there exists an image towards the direction (represented by
an arrow) parallel to that tangent from the image plane centre. Note
that the image plane centre is not necessarily coincident with the
source plane centre nor with the caustics centre. The magnification of the
image is proportional to the inverse of the distance between the
source and the tangent point.
}\label{ims}
\end{figure}

So far, the discussion on the LCIP pair has been focused on the
identical (up to linear approximation) caustics and the degenerate
(total) magnification. A natural question to follow is whether they are
indistinguishable (at least up to the linear order) in every sense.
The answer is actually negative as one can easily discover the
difference in the critical curves and the individual image positions.
For the monomial perturbation of $f(z)=az^n$, one finds, from
equation~(\ref{lcr}), the corresponding critical curves to be;
\begin{equation}
\label{pcr}
z_c(\phi)=\E^{\I\phi}\left\{
1-\frac{\epsilon}{2}n|a|\cos[(n+1)\phi+\phi_a]\right\},
\end{equation}
where $\phi_a$ is the argument of $a$, that is, $a=|a|\E^{\I\phi_a}$.
It is straightforward to show that equation~(\ref{pcr}) is \emph{not}
invariant under the LCIP transformation of $n\leftrightarrow-(n+2)$
and $a\leftrightarrow-\ab$ (equivalently, $\phi_a\leftrightarrow
\upi-\phi_a$), but rather the amplitude of the cosine linear deviation
term changes to $\epsilon(n+2)|a|/2$. In addition, while the azimuthal
positions of the images for the LCIP pair are the same since
equation~(\ref{peq}) is identical for them, one finds that the radial
positional deviations given by
\begin{equation}
\label{rad}
\delta r=\frac{\E^{-\I\phi}}{2}\left[
\zeta+\epsilon\f\!(\E^{-\I\phi})\right]\ (\in\mathbf R)
\end{equation}
cannot be the same for the two different perturbations. However, it
should be noted that the leading order descriptions of the critical
curves and the image radial positions are dominated not by the first
order but by the zeroth order terms, and they are still identical up
to their leading zeroth order. More importantly, despite this
difference, the actual observational distinction between the LCIP pair
would be rather difficult since the critical curve is physically
unobservable and two dominant observable characteristics of the
individual images -- the magnification and the azimuthal position --
are indeed identical up to the linear order of the perturbations.

To find the magnification and the azimuthal position of the individual
images, one must solve equation~(\ref{peq}), which is a
transcendental equation for $\phi$, and thus it may not be possible to
find algebraic solutions although it is straightforward to find its
roots graphically. Incidentally, it is also possible to find them in
a purely geometrical manner provided that one can draw its caustics
(eq.~\ref{cau}). From equation~(\ref{dzp}), one finds that
$(d\zeta_c/d\phi)\parallel \E^{\I\phi}$, that is, the parameter $\phi$
also determines the tangent direction to the caustics. This, together
with the fact that equation~(\ref{ljc}) is real when $\phi$ is the
solution of equation~(\ref{peq}), also implies that, if one can draw a
tangent to the caustic through the given source position, the angle of
that tangent line $\phi$ is a root of equation~(\ref{peq}). Hence,
there exists an image with its azimuthal position with respect to the
image plane centre being $\phi$ and magnification given by
$(2l)^{-1}$ where $l$ is the distance between the source and the
tangent point. Examples of this procedure for the perturbation given
by $f(z)=az$ or its LCIP $f(z)=-\ab z^{-2}$ is found in
Fig.~\ref{ims}. For definiteness, $a$ is assumed to be real
(horizontal direction) in this example, and therefore, the two cusps lying
along the horizontal direction are positive. Finally, note that not
all images are actually solutions of
equation~(\ref{peq}). However, missing images, if any, are typically
higher order solutions that do not significantly contribute to the
total magnification provided that the linear approximation is valid
and dominant and that the source lies sufficiently close to the centre (and
therefore the caustics).

\subsection{What is so special about the point-mass lens?}
\label{ssec_csl}
The discussion until now has been limited to the case for which the
unperturbed lens system is described by the point-mass lens
equation~(\ref{le0}). However, the perturbation approach outlined in
Section~\ref{sec_per} can be applied to any lens system whose solutions are
known. In fact, the procedure in Section~\ref{sec_per} can be generalized
to the case for which the unperturbed lens system is given by any
circularly symmetric lens system that can form an Einstein ring. Suppose
that the system is described by the lens equation~(\ref{ple}) with the
unperturbed part given by
\begin{equation}
\zeta_s=z\left(1-\frac{1}{r}\frac{d\psi}{dr}\right)
=z\left[1-\frac{4G}{D_\mathrm{rel}c^2}\frac{M(<r)}{r^2}\right],
\end{equation}
where $r=|z|$ is the radial coordinate of image, $\psi(r)$ is a
circularly symmetric lensing potential, and $M(<r)$ is the mass
enclosed within a given angular radius $r$. If the equation
\begin{equation}
\label{ein}
\frac{4G}{D_\mathrm{rel}c^2}M(<r)=r^2
\end{equation}
possesses a positive root $r=r_0$, one can find the linear
approximation for the tangential caustics of the system by
series-expanding the lens equation at $z_0=r_0\E^{\I\phi}$ and
following procedure similar to Section~\ref{ssec_ccc}. In particular,
one finds
\begin{equation}
\zeta=(2-\beta_0)\,\delta r \E^{\I\phi}-\epsilon\f\!(r_0\E^{-\I\phi}),
\end{equation}
\begin{equation}
F\!(\phi)=\zeta+\epsilon\f\!(r_0\E^{-\I\phi})-
\E^{2\I\phi}[\bar\zeta+\epsilon f\!(r_0\E^{\I\phi})]=0,
\end{equation}
\begin{equation}
\J=(2-\beta_0)\,\frac{\zeta-\zeta_c(\phi)}{z_0},
\end{equation}
\begin{equation}
\frac{\zeta_c(\phi)}{r_0}=-\frac{\epsilon}{2}
\left[\E^{3\I\phi}f^\prime\!(r_0\E^{\I\phi})+
\E^{-\I\phi}\f^\prime\!(r_0\E^{-\I\phi})+
2\frac{\f\!(r_0\E^{-\I\phi})}{r_0}\right],
\end{equation}
and
\begin{equation}
\frac{z_c(\phi)}{r_0}=
\E^{\I\phi}-\frac{\epsilon}{2(2-\beta_0)}
\left[\E^{3\I\phi}f^\prime\!(r_0\E^{\I\phi})+
\E^{-\I\phi}\f^\prime\!(r_0\E^{-\I\phi})\right],
\end{equation}
in place of equations~(\ref{lle}), (\ref{peq}), (\ref{ljc}),
(\ref{cau}), and (\ref{lcr}). Here, $\beta_0=(d\ln M/d\ln r)|_{r=r_0}$
is the local power index for the mass $M$ at the projected angular
radius $r=r_0$. In other words, the shape of (the linear approximation
of) the caustics is completely determined by the perturbation alone
once they are scaled by the angular Einstein ring radius (note that
eq.~\ref{ein} actually defines $r_0$ to be the angular Einstein ring
radius for the given circularly symmetric system), and the unperturbed
mass distribution only affects the image magnification as a common
multiplicative factor, and therefore, the magnification ratios between
images are also completely independent of the unperturbed mass
distribution. Moreover, it is easy to establish the same LCIP relation
$f(z)=ar_0(z/r_0)^n$ and $f(z)=-\ab r_0(z/r_0)^{-(n+2)}$, or more
generally, that the pair of perturbation potentials with an inverse
symmetry with respect to Einstein ring radius $r_0$ [i.e.,
$\delta\psi^t(\bmath r)$ and $\delta\psi^p(\bmath r)=
\delta\psi^t(r_0^2\bmath r/r^2)$] has the same expression for the
linear approximation of the tangential caustics.

\subsection{Are there LCIP pairs for convergent perturbations?}
\label{ssec_scp}
As has been noted, the LCIP relation originates from a certain
intrinsic symmetry of harmonic functions, that is, their azimuthal
structures are invariant under the radial distance inversion. This
implies that the two-fold correspondence LCIP does not exist for
convergent perturbations, for which the perturbation potential is no
longer harmonic. However, examination of equations~(\ref{pty}) and
(\ref{pmp}) reveals more basic properties of the LCIP, that is, the
azimuthal Fourier coefficients of the perturbation potentials at the
angular Einstein ring radius are the same, which is indeed the
generalization of LCIP to include convergent perturbations.

Let us think of the potential
\begin{equation}
\label{ppo}
\delta\psi(r,\phi)=
\sum_n\left[\epsilon_n(r)\cos n\phi+\varrho_n(r)\sin n\phi\right]=
\frac{1}{2}\sum_n\left[\frac{z^n}{r^n}\bar\varepsilon_n(r)+
\frac{r^n}{z^n}\varepsilon_n(r)\right]
\end{equation}
perturbing a circularly symmetric lens system. Here, $\varepsilon_n=
\epsilon_n+\I\varrho_n$. Then, the lens equation is given by
\begin{equation}
\label{gle}
\zeta=z\left(1-\frac{4G}{D_\mathrm{rel}c^2}\frac{M}{r^2}\right)-\delta\alpha,
\end{equation}
where
\begin{equation}
\label{dap}
\delta\alpha=2\upartial_{\z}\delta\psi=\sum_n
\left[\frac{z^{n+1}}{2r}\frac{d}{dr}\left(\frac{\bar\varepsilon_n}{r^n}\right)
+\frac{1}{z^{n-1}}\frac{1}{2r}\frac{d}{dr}\left(r^n\varepsilon_n\right)\right]
=\frac{r_0}{2}\sum_n
\left[\frac{r_0^{n+1}}{\z^{n+1}}\bar\fz_n
-\frac{\z^{n-1}}{r_0^{n-1}}\gz_n\right].
\end{equation}
\begin{equation}
\fz_n\!(r)=\frac{r^{2n+1}}{r_0^{n+2}}
\frac{d}{dr}\left(\frac{\varepsilon_n}{r^n}\right)
\,;\ \ \
\gz_n\!(r)=-\frac{r_0^{n-2}}{r^{2n-1}}
\frac{d}{dr}\left(r^n\varepsilon_n\right)
\end{equation}
Here, the calculation utilizes the fact that $r^2=z\z$ and
consequently that $\upartial_{\z} r=z(2r)^{-1}=r(2\z)^{-1}$. That is, for
any real function $F(r)$, one can relate the complex derivative
and the ordinary derivatives via $\upartial_{\z} F=F'\upartial_{\z} r=zF'
(2r)^{-1}=rF'(2\z)^{-1}$. To find the expression for the linear
approximation of the caustics, one series-expands equation~(\ref{gle})
and its Jacobian determinant at $z_0=r_0\E^{\I\phi}$ as in
Section~\ref{ssec_ccc}. After some tedious but none the less straightforward
algebra, one can find
\begin{equation}
\label{cvj}
\J=2(1-\kappa_{s,0})\,\frac{\zeta-\zeta_c(\phi)}{z_0}
\end{equation}
\begin{equation}
\label{cvc}
\zeta_c(\phi)=\frac{r_0}{4}\sum_n\left[
(n+1)\left(\fz_{n,0}+\gz_{n,0}\right)\E^{-(n-1)\I\phi}+
(n-1)\overline{\left(\fz_{n,0}+\gz_{n,0}\right)}\E^{(n+1)\I\phi}
\right]
\end{equation}
in place of equations~(\ref{ljc}) and (\ref{cau}), where
$\kappa_{s,0}$ is the (local) convergence of the unperturbed
circularly symmetric lens at $r=r_0$ (note that $\beta_0=
2\kappa_{s,0}$ at $r=r_0$), and the subscript `$_0$' indicates the
value of the function at $r=r_0$ [e.g., $\fz_{n,0}=
\fz_n(r_0)$ and $\gz_{n,0}=\gz_n(r_0)$]. Then,
\begin{equation}
\fz_{n,0}=\fz_n\!(r_0)=r_0^{n-1}
\left.\frac{d}{dr}\left(\frac{\varepsilon_n}{r^n}\right)\right|_{r=r_0}=
\frac{\varepsilon^\prime_{n,0}}{r_0}-n\frac{\varepsilon_{n,0}}{r_0^2}
\end{equation}
\begin{equation}
\gz_{n,0}=\gz_n\!(r_0)=-\frac{1}{r_0^{n+1}}
\left.\frac{d}{dr}\left(r^n\varepsilon_n\right)\right|_{r=r_0}=
-\frac{\varepsilon^\prime_{n,0}}{r_0}-n\frac{\varepsilon_{n,0}}{r_0^2}
\end{equation}
so that
\begin{equation}
\zeta_c(\phi)=-\frac{r_0}{2}\sum_nn\left[
(n+1)\E^{-(n-1)\I\phi}\frac{\varepsilon_{n,0}}{r_0^2}+
(n-1)\E^{(n+1)\I\phi}\frac{\bar\varepsilon_{n,0}}{r_0^2}\right],
\end{equation}
or
\begin{equation}
\frac{\Re[\zeta_c]}{r_0}=-\sum_nn\left[
\frac{\epsilon_{n,0}}{r_0^2}(n\cos n\phi\cos\phi+\sin n\phi\sin\phi)+
\frac{\varrho_{n,0}}{r_0^2}(n\sin n\phi\cos\phi-\cos n\phi\sin\phi)\right]
\end{equation}
\begin{equation}
\frac{\Im[\zeta_c]}{r_0}=-\sum_nn\left[
\frac{\epsilon_{n,0}}{r_0^2}(n\cos n\phi\sin\phi-\sin n\phi\cos\phi)+
\frac{\varrho_{n,0}}{r_0^2}(n\sin n\phi\sin\phi+\cos n\phi\cos\phi)\right].
\end{equation}
Note that the expressions do not involve the radial derivatives at
all. In other words, up to the linear order, the caustics are
completely specified by $\varepsilon_{n,0}=\epsilon_n(r_0)+
\I\varrho_n(r_0)$ or the azimuthal behaviour of the potential
$\delta\psi(r_0,\phi)$ exactly \emph{at} $r=r_0$, and independent of
its radial behaviour even in its immediate neighbourhood. In terms of
the deflections due to the perturbations (instead of the perturbation
potentials) given as in equation~(\ref{dap}), this translates into
the linear approximation of the caustics being a function of
$\fz_{n,0}+\gz_{n,0}$ alone, and thus the two
perturbations producing the same linear approximation of the
caustics if $\fz_{n,0}+\gz_{n,0}$ is the same. The
null-convergent LCIP pair is a special case of a pair of two
perturbations given respectively by
$(\fz_n,\gz_n)=(c_n,0)$ and
$(\fz_n,\gz_n)=(0,c_n)$ where the $c_n$'s are (complex)
constants. In terms of the potentials, they are the pair of
$\varepsilon^t_n=a_n(r/r_0)^n$ and $\varepsilon^p_n=a_n(r_0/r)^n$
where the $a_n$'s are again some common constants.

\subsubsection{ellipticity, external shear, and quadrupole moment}
In many applications of gravitational lensing, the lensing
convergence may be elliptically symmetric, that is, $\kappa(r,\phi)=
\kappa_a[a(r,\phi)]$ is the function of $r$ and $\phi$ only through
the combination $a=r\rho^{-1}(\sin^2\phi+\rho^2\cos^2\phi)^{1/2}$,
where $\rho$
is a constant. It is straightforward to show that this yields
concentric ellipses with eccentricity of $|1-\rho^2|^{1/2}$ as
iso-convergence contours. The analysis of these types of lensing systems
is of great interest, but also very much more complicated than the
circularly symmetric system. However, if $\rho\simeq1$, the problem can
be greatly simplified by applying Fourier analysis. Let us consider
the radial Taylor expansion of $\kappa(r,\phi)$ with fixed $\phi$,
that is, $\kappa(r,\phi)\approx\kappa(\tilde r,\phi)+(d\kappa/dr)
|_{r=\tilde r} (r-\tilde r)$. Suppose that the expansion was at
$\tilde r=2^{1/2}a (1+\rho^{-2})^{-1/2}$. Then the convergence at
$r=a\rho(\sin^2\phi+\rho^2\cos^2\phi)^{-1/2}$ is
\begin{equation}
\label{ell}
\kappa_a(a)\approx\kappa(\tilde r,\phi)\left\{1+
\left.\frac{d\ln\kappa}{d\ln r}\right|_{r=\tilde r}
\left[\left(1-\frac{1-\rho^2}{1+\rho^2}\cos2\phi\right)^{-1/2}
-1\right]\right\}.
\end{equation}
Here, by noticing that $\kappa_a(a)$ and $(d\ln\kappa/d\ln r)|_{r=
\tilde r}$ do not involve any azimuthal dependence,
equation~(\ref{ell}) can be used to derive the quadrupole approximation
\citep[c.f.,][]{Ko87a}
of the elliptically symmetric convergence
\begin{equation}
\kappa(\tilde r,\phi)\approx
\kappa_a\left[\frac{\tilde r}{\rho}\sqrt{\frac{1+\rho^2}{2}}\right]
\left(1-\frac{1}{2}\,\frac{1-\rho^2}{1+\rho^2}
\left.\frac{d\ln\kappa}{d\ln r}\right|_{r=\tilde r}
\cos2\phi\right),
\end{equation}
provided that $|1-\rho^2|\ll1$.

Suppose that the lensing convergence is given by
\begin{equation}
\label{con}
\kappa(r,\phi)=\kappa_s(r)
\left\{1+\xi\cos[2(\phi-\phi_e)]\right\}.
\end{equation}
Then, with the canonical boundary condition, the potential can be
found, using Green's function, to be
\begin{equation}
\psi(r,\phi)=2\left[
\ln r\int_0^r\!\D\hat r\,\hat r\kappa_s(\hat r)+
\int_r^\infty\!\D\hat r\,\hat r\kappa_s(\hat r)\ln\hat r\right]
-\frac{1}{2}\left\{
\frac{1}{r^2}\int_0^r\!\D\hat r\,\hat r^3\xi\kappa_s(\hat r)
\cos[2(\phi-\phi_e)]+r^2\int_r^\infty\!\frac{\D\hat r}{\hat r}\,
\xi\kappa_s(\hat r)\cos[2(\phi-\phi_e)]\right\}.
\end{equation}
If one defines radial functions $\mu_0(r)\equiv\int_0^r\kappa_s
\hat r\D\hat r$, which is basically a scaled mass associated with the
circularly symmetric part of the convergence, and $\mu_1(r)\equiv
\int_0^r\xi\kappa_s\hat r\D\hat r$, then the potential (up to an
addictive constant) may also be written in (assuming $\phi_e$ is
constant)
\begin{equation}
\label{pot}
\psi(r,\phi)=2\psi_0+
2\int_{\underline{r}}^r\!\D\hat r\,\frac{\mu_0(\hat r)}{\hat r}
-\frac{1}{2}\left[
\frac{1}{r^2}\int_0^r\!\D\hat r\,\hat r^2\mu_1^\prime(\hat r)
+r^2\int_r^\infty\!\D\hat r\,
\frac{\mu_1^\prime(\hat r)}{\hat r^2}\right]\cos[2(\phi-\phi_e)],
\end{equation}
where $\underline{r}$ is some fiducial radius, $\psi_0$ is a constant
that may be formally defined by $\psi_0\equiv\mu_0(\underline{r})
\ln\underline{r}+\int_{\underline{r}}^\infty\mu_0'(r)\ln r\D r$. In
addition, one can also find the deflection function
\begin{equation}
\label{def}
\alpha=2\upartial_{\z}\psi=2\frac{\mu_0(r)}{r^2}z+
\frac{1}{\z^3}\int_0^r\!\D\hat r\,\hat r^2\mu_1^\prime(\hat r)\E^{-2\I\phi_e}
-\z\int_r^\infty\!\D\hat r\,\frac{\mu_1^\prime(\hat r)}{\hat r^2}
\E^{2\I\phi_e}.
\end{equation}
One can note that the potential given by equation~(\ref{pot}) can be
considered as a circularly symmetric potential perturbed by a small
perturbing potential of the form of equation~(\ref{ppo}) with $n=2$
provided that both of the integrals are small. Assuming that the
circularly symmetric part of the potential in equation~(\ref{pot})
allows to form an Einstein ring at $r=r_0$ [i.e., $2\mu_0(r_0)=
r_0^2$], the linear approximation of the tangential caustics is given
by
\begin{equation}
\label{cca}
\frac{\zeta_c(\phi)}{r_0\E^{\I\phi_e}}=
\frac{(\fz^\star_0+\gz^\star_0)}{2}
\left[3\E^{-\I(\phi-\phi_e)}+\E^{3\I(\phi-\phi_e)}\right]
=2(\fz^\star_0+\gz^\star_0)
\left[\cos^3(\phi-\phi_e)-\I\sin^3(\phi-\phi_e)\right]
\end{equation}
\begin{equation}
\fz^\star_0=\int_0^{r_0}\!\frac{\D r}{r_0}
\left(\frac{r}{r_0}\right)^2
\frac{\mu_1^\prime(r)}{r_0}
\,;\ \ \
\gz^\star_0=\int_{r_0}^\infty\!\frac{\D r}{r_0}
\left(\frac{r_0}{r}\right)^2
\frac{\mu_1^\prime(r)}{r_0}.
\end{equation}
Here, the caustic is a tetracuspoid so that there are four images
for a source position within the caustic. Equation~(\ref{cca}),
from a comparison to equations~(\ref{ccc}), (\ref{cri}), and
(\ref{are}), also implies that the cross-section of quadruple images
is $(3\upi/2)(\fz^\star_0+ \gz^\star_0)^2$. Furthermore,
the lensing behaviour of the source near the lens centre
is essentially described by two \emph{constants}
$\fz^\star_0$ and $\gz^\star_0$. In particular,
when the source and the centre of the lens are perfectly aligned
($\zeta=0$), there are four images associated with the caustics, and
their azimuthal coordinates are given by $\phi=n\upi/2+\phi_e$ where
$n$ is an integer. It is easy to verify that equation~(\ref{cvj}) is
real and that
\begin{equation}
z=(r_0+\delta r)\E^{\I\phi}=
r_0\E^{\I\phi_e}\left[1+(-1)^n
\frac{(\fz^\star_0-\gz^\star_0)}{2(1-\kappa_{s,0})}
\right]\E^{n\I\upi/2}
\,;\ \ \
\J=(-1)^{n+1}4(1-\kappa_{s,0})
(\fz^\star_0+ \gz^\star_0)
\end{equation}
so that the resulting images are an alternating-parity
equal-magnification two-fold symmetric crucifix-form quartet with an
axis ratio of $1-|(\fz^\star_0-\gz^\star_0)/(1-\kappa_{s,0})|$.
For more general source positions,
once the caustic is drawn using equation~(\ref{cca}), the azimuthal
coordinates and the individual magnifications of the images can be found
by following the same procedure as in Section~\ref{ssec_wcc} using
equation~(\ref{cvj}). In addition, since the form of the perturbation
here is specified, it is also possible to find the radial coordinates
of the images by the equivalent of equation~(\ref{rad})
\begin{equation}
\label{dtr}
\delta r=
\frac{\E^{-\I\phi}}{2(1-\kappa_{s,0})}\left\{\zeta+r_0\left[
\E^{\I(3\phi-2\phi_e)}\fz^\star_0
-\E^{\I(2\phi_e-\phi)}\gz^\star_0
\right]\right\}=\frac{\zeta \E^{-\I\phi}+r_0\left[
(\fz^\star_0-\gz^\star_0)\cos[2(\phi-\phi_e)]
+\I(\fz^\star_0+\gz^\star_0)\sin[2(\phi-\phi_e)]
\right\}}{2(1-\kappa_{s,0})},
\end{equation}
for the given radial coordinate of the images $\phi$. Algebraically,
$\phi$ can be determined from the condition that $\delta r$ is real,
that is, $\Im[\zeta \E^{-\I\phi}]+(\fz^\star_0+
\gz^\star_0)\sin[2(\phi-\phi_e)]=0$. In other words, for a
given source position, $\phi$ is determined by $\fz^\star_0
+\gz^\star_0$, which is in accordance with the geometric
construction of $\phi$ since it is the same coefficient for the
caustics (eq.~\ref{cca}).

Note the linear approximation of the caustics (eq.~\ref{cca})
is identical (after a
proper rotation of coordinate) to those caused either by the
null-convergent compact quadrupole perturbation with the quadrupole
moment of $(\fz^\star_0+\gz^\star_0)r_0^4$ or by
the null-convergent tidal perturbation by a constant external shear
of $-(\fz^\star_0+\gz^\star_0)$
\citep*[c.f.,][]{WMS95}.
In fact, the examination of
equation~(\ref{def}) suggests that those locally (near Einstein ring
region) null-convergent perturbations can be caused by the quadrupole
moment of the mass distribution, either entirely enclosed within the
angular Einstein ring radius ($f\propto z^{-3}$), or lying completely
outside of the angular Einstein ring radius ($f\propto z$). In
general, the lensing behaviour for the source near the lens centre when
the convergence is given by equation~(\ref{con}) with $|\xi|\ll1$
is basically equivalent to that of the system with a circularly
symmetric convergence $\kappa_s$ under the perturbation of $\f=r_0
[\bar\fz^\dag_0(\z/r_0)^{-3}-\gz^\dag_0(\z/r_0)]$, where
$\fz^\dag_0=\fz^\star_0\E^{2\I\phi_e}$ and $\gz^\dag_0=
\gz^\star_0\E^{2\I\phi_e}$. This can be generalized to the
elliptically symmetric
system that is additionally affected by external shear and/or compact
quadrupole mass moment.\footnote{In much of literature, a quadrupole lens
and a lens with shear are used interchangeably without distinction.
Although this practice has a certain justification since both the effects
due to the constant external shear and the compact quadrupole moment
are associated with $\cos2\phi$ terms in the azimuthal part of the
potential, they are always explicitly distinguished in this paper.}
For this case, $\fz^\dag_0=
\fz^\star_0\E^{2\I\phi_e}+\hat Q_4$ and $\gz^\dag_0=
\gz^\star_0\E^{2\I\phi_e}+\gamma_0$,\footnote{The sign for
the external shear is chosen to follow the usual convention in
strong lensing literatures \citep[e.g.,][]{HKM04}. This
choice leads the lensing shear due to the external shear term to be
`$-\gamma_0$,' as defined in Appendix~\ref{asec_ceq}.} and they are
not necessarily parallel to each other any more. Here, note that
$\hat Q_4=|\hat Q_4|\E^{2\I\phi_Q}$ and $\gamma_0=|\gamma_0|
\E^{2\I\phi_\gamma}$ are complex numbers. (Strictly speaking, they are
complex number representations of certain symmetric 2$\times$2 traceless
tensors. The absolute value of the complex number corresponds to the
positive eigenvalue of the tensor and its argument is the same as twice
the angle between the corresponding eigenvector and the real direction.)
Since equations~(\ref{cvj}) and (\ref{cvc}) are still valid regardless of
$\fz_n$ and $\gz_n$ not being parallel, one finds, in
the place of equation~(\ref{cca}), -- note the factor of two
difference in the definition compared to eq.~(\ref{dap}) -- that
\begin{equation}
\zeta_c(\phi)=\frac{r_0\E^{\I\phi_+}}{2}
\left|\fz^\dag_0+\gz^\dag_0\right|
\left[3\E^{-\I(\phi-\phi_+)}+\E^{3\I(\phi-\phi_+)}\right]
=2r_0\E^{\I\phi_+}\left|\fz^\dag_0+\gz^\dag_0\right|
\left[\cos^3(\phi-\phi_+)-\I\sin^3(\phi-\phi_+)\right]
\end{equation}
and also that
\begin{equation}
\delta r=
\frac{\E^{-\I\phi}}{2(1-\kappa_{s,0})}\left\{\zeta+r_0\left[
\E^{3\I\phi}\bar\fz^\dag_0-\E^{-\I\phi}\gz^\dag_0
\right]\right\}=\frac{\zeta \E^{-\I\phi}+r_0\left\{
\left|\fz^\dag_0-\gz^\dag_0\right|\cos[2(\phi-\phi_-)]
+\I\left|\fz^\dag_0+\gz^\dag_0\right|\sin[2(\phi-\phi_+)]
\right\}}{2(1-\kappa_{s,0})},
\end{equation}
in place of equation~(\ref{dtr}). Here, $\fz^\dag_0+
\gz^\dag_0=|\fz^\dag_0+\gz^\dag_0|\E^{2\I\phi_+}$ and
$\fz^\dag_0-\gz^\dag_0=|\fz^\dag_0-\gz^\dag_0|
\E^{2\I\phi_-}$. As before, $\delta r\in\mathbf R$ for $\phi$
that is the azimuthal coordinate of the image for the given source
position. For $\zeta=0$, one finds the azimuthal coordinates of images
$\phi=n\upi/2+\phi_+$ and also that $2(1-\kappa_{s,0})\delta r=(-1)^n
r_0|\fz^\dag_0-\gz^\dag_0|\cos[2(\phi_+-\phi_-)]$. This implies
that the resulting image quartets are identical for the set of
$\fz^\dag_0$ and $\gz^\dag_0$ if $|\fz^\dag_0+\gz^\dag_0|$
and $|\fz^\dag_0-\gz^\dag_0||\cos[2(\phi_+-\phi_-)]|$ are
constant. For a more general source position $\zeta\ne0$, while one
finds that the azimuthal coordinates and the magnification ratios of
images are completely specified by $\fz^\dag_0+\gz^\dag_0$, the
projection that controls the radial coordinates depends on each
azimuthal coordinate $\phi$ of the image, so that they are in general not
common for the four images. However, if the source is sufficiently close
to the centre,
the projections
are more or less parallel to one another, and therefore, there still
exists a certain near degeneracy of the images concerning the choice
of $|\fz^\dag_0|$, $|\gz^\dag_0|$, and their relative orientation
in the underlying lens model.

If the convergence is given by a power-law ellipsoid
\begin{equation}
\kappa[a(r,\phi)]=\frac{\beta}{2}\left(\frac{a_0}{a}\right)^{2-\beta}=
\frac{\beta}{2}\left(\frac{a_0}{r}\right)^{2-\beta}
\left(\rho^{-2}\sin^2\phi+\cos^2\phi\right)^{-(2-\beta)/2}=
\frac{\beta}{2}\left(\frac{a_0}{r}\right)^{2-\beta}
\left(\frac{2\rho^2}{1+\rho^2}\right)^{(2-\beta)/2}
\left(1-\frac{1-\rho^2}{1+\rho^2}\cos2\phi\right)^{-(2-\beta)/2},
\end{equation}
it can be approximated by a quadrupole convergence in a form of
equation~(\ref{con}) with
\begin{equation}
\kappa_s=\frac{\beta}{2}\left(\frac{a_0}{r}\right)^{2-\beta}
\left(\frac{2\rho^2}{1+\rho^2}\right)^{(2-\beta)/2}\,;\ \ \
\xi=\frac{1-\rho^2}{1+\rho^2}\left(1-\frac{\beta}{2}\right).
\end{equation}
Here, $0<\beta<2$. In addition,
\begin{equation}
\mu_0(r)=\int_0^r\!\D\hat r\,\hat r\kappa_s(\hat r)=
\frac{\beta}{2}\left(\frac{2\rho^2}{1+\rho^2}\right)^{(2-\beta)/2}
a_0^{2-\beta}\int_0^r\!\D\hat r\,\hat r^{\beta-1}=
\left(\frac{2\rho^2}{1+\rho^2}\right)^{(2-\beta)/2}
\frac{a_0^2}{2}\left(\frac{r}{a_0}\right)^\beta
\end{equation}
so that one finds $r_0=2^{1/2}a_0(1+\rho^{-2})^{-1/2}$ by solving
$2\mu_0(r_0)=r_0^2$. Furthermore, $\mu_1(r)=\xi\mu_0(r)$ since
$\xi$ is now constant, and $2\kappa_{s,0}=\beta$. Note that
$2\mu_0(r)=r_0^{2-\beta}r^\beta$ and $(d\ln\mu_0/d\ln r)=
(d\ln\mu_1/d\ln r)=\beta$. Finally,
\begin{equation}
\fz^\star_0=\int_0^{r_0}\!\frac{\D r}{r_0}
\left(\frac{r}{r_0}\right)^2\frac{\mu_1^\prime(r)}{r_0}=
\xi\frac{\beta}{2}\int_0^1\!\D\rdash\,\rdash^{\beta+1}=
\frac{\xi\beta}{2(\beta+2)}=
\frac{1-\rho^2}{1+\rho^2}\frac{\beta}{4}\frac{2-\beta}{2+\beta},
\end{equation}
\begin{equation}
\gz^\star_0=\int_{r_0}^\infty\!\frac{\D r}{r_0}
\left(\frac{r_0}{r}\right)^2\frac{\mu_1^\prime(r)}{r_0}=
\xi\frac{\beta}{2}\int_1^\infty\!\D\rdash\,\rdash^{\beta-3}=
-\frac{\xi\beta}{2(\beta-2)}=
\frac{1-\rho^2}{1+\rho^2}\frac{\beta}{4},
\end{equation}
and thus
\begin{equation}
\fz^\star_0+\gz^\star_0=
\frac{1-\rho^2}{1+\rho^2}\frac{\beta}{2+\beta}
\,;\ \ \
\fz^\star_0-\gz^\star_0=
-\frac{1-\rho^2}{1+\rho^2}\frac{\beta^2}{2(2+\beta)}.
\end{equation}
Hence, for a singular isothermal ellipsoid (SIE) lens
($2\kappa_{s,0}=\beta=1$), with an
ellipticity parameter $\ee=(1-\rho^2)/(1+\rho^2)\simeq
1-\rho$, one finds that
$\fz^\star_0+\gz^\star_0=\ee/3$ and
$\fz^\star_0-\gz^\star_0=-\ee/6$, and thus, the
four-image cross-section is given by 
$(3\upi/2)(\fz^\star_0+\gz^\star_0)^2=(\upi/6)\ee^2$,
and the axis ratio for a crucifix-form quartet is 
$1-|(\fz^\star_0-\gz^\star_0)/(1-\kappa_{s,0})|=1-(\ee/3)$
\citep*{KKS97}.

If this SIE lens is subject to perturbations due to the constant external
shear, one finds that
\begin{equation}
\fz_0^\dag=
\frac{\ee}{12}\E^{2\I\phi_e}=
\frac{1}{12}\left(\ee\E^{2\I\phi_e}+4|\gamma_0|e^{2\I\phi_\gamma}\right)
-\frac{|\gamma_0|}{3}\E^{2\I\phi_\gamma}\,;\ \ \
\gz_0^\dag=
\frac{\ee}{4}\E^{2\I\phi_e}+|\gamma_0|\E^{2\I\phi_\gamma}=
\frac{1}{4}\left(\ee\E^{2\I\phi_e}+4|\gamma_0|e^{2\I\phi_\gamma}\right).
\end{equation}
By comparing this to another SIE lens system that is affected by the
compact quadrupole moment,
$\fz_0^\dag=
(\tilde\ee/12)\E^{2\I\tilde\phi_e}+|\hat Q_4|\E^{2\I\phi_Q}$ and
$\gz_0^\dag=(\tilde\ee/4)\E^{2\I\tilde\phi_e}$,
one can conclude that the lensing behaviours of two systems are identical
up to the linear approximation provided that the external shear and
the compact quadrupole moment are related by $|\gamma_0|=3|\hat Q_4|$
and $\E^{2\I(\phi_Q-\phi_\gamma)}=-1$ and the two ellipticity parameters
by
\begin{equation}
\tilde\ee\E^{2\I(\tilde\phi_e-\phi_e)}=\ee\left[1+
\frac{4|\gamma_0|}{\ee}\E^{2\I(\phi_\gamma-\phi_e)}\right]
\,;\ \ \
\ee\E^{2\I(\phi_e-\tilde\phi_e)}=\tilde\ee\left[1+
\frac{12|\hat Q_4|}{\tilde\ee}\E^{2\I(\phi_Q-\tilde\phi_e)}\right].
\end{equation}
In other words, the lens system that can be modelled by a SIE lens with
an external shear can also be modelled by a different SIE lens with a
quadrupole moment, the magnitude of which is a third of
that of the external shear,
provided that the system's departure from the circular symmetry is small.
This can be further generalized to systems that are subject to an external
shear and a compact quadrupole moment at the same time to yield a certain
degeneracy of the lens model regarding the ellipticity, the shear, and
the quadrupole moment (as well as their relative orientations).

If one models an image quartet using a SIE lens with a constant
external shear but no quadrupole moment, then
$\fz_0^\dag+\gz_0^\dag=|\fz_0^\dag+\gz_0^\dag|\E^{2\I\phi_+}=
(\ee/3)\E^{2\I\phi_e}+|\gamma_0|\E^{2\I\phi_\gamma}$ and
$\fz_0^\dag-\gz_0^\dag=|\fz_0^\dag-\gz_0^\dag|\E^{2\I\phi_-}=
-(\ee/6)\E^{2\I\phi_e}-|\gamma_0|\E^{2\I\phi_\gamma}$.
Hence, the four-image cross-section is found to be
\begin{equation}
\frac{3\upi}{2}\left|\fz_0^\dag+\gz_0^\dag\right|^2=
\upi\left\{\frac{\ee^2}{6}+\frac{3|\gamma_0|^2}{2}+
\ee|\gamma_0|\cos[2(\phi_\gamma-\phi_e)]\right\}.
\end{equation}
It is easy to see that, for given $\ee$ and $|\gamma_0|$,
the cross-section is largest [$=\upi(\ee+3|\gamma_0|)^2/6$]
if $\phi_\gamma=\phi_e$ while it is at minimum
[$=\upi(\ee-3|\gamma_0|)^2/6$] when $\phi_\gamma-\phi_e=\upi/2$.
In addition, one can find that
\begin{equation}
\left(\frac{\ee}{3}\right)^2-|\gamma_0|^2=
|\fz_0^\dag+\gz_0^\dag|\left\{3|\fz_0^\dag+\gz_0^\dag|
+4|\fz_0^\dag-\gz_0^\dag|\cos[2(\phi_+-\phi_-)]\right\}.
\end{equation}
That is to say, for fixed $|\fz_0^\dag+\gz_0^\dag|$ and
$|\fz_0^\dag-\gz_0^\dag|\cos[2(\phi_+-\phi_-)]$, which produce
identical crucifix-form image quartets, there exists a degeneracy running
along a hyperbolic path in ($\ee$, $|\gamma_0|$) space. This may be
seen as a prototype for the `cancellation branches' of the `{\sf U}' shape
degenerate path found in fig.~8 of \citet{KKS97}. In reality, it is
highly unlikely that the source is perfectly aligned with the lens centre
so that the actual degeneracy becomes rather more complex.
(Furthermore, as the ellipticity and/or the shear get larger,
the whole perturbative approach will start to break down.) Nevertheless,
the basic idea that there exists a certain set of 
$\fz_0^\dag+\gz_0^\dag$ and $\fz_0^\dag-\gz_0^\dag$ that yields
similar (or nearly degenerate) image configurations (with the proper
rotation of frames) is thought to be more or less valid, and this can
lead to the observed degeneracy of the shear and ellipticity
combination.

\section*{acknowledgment}
The author thanks N.~Wyn Evans for pointing out possible connection
of the subject to macrolens modelling.
The author is grateful to A.~Gould for his careful reading of the
manuscript.
Fig.~\ref{alc} is reproduced from \citet{Al02}, in a modified form,
by permission of the American Astronomical Society.

\appendix

\section{complex analytic function}
\label{asec_caf}

Let $f=u(x,y)+\I v(x,y)$ be a function of $z=x+\I y$ in the region 
$R\subset\mathbf C$
containing $z_0$, where $u$ and $v$ are real-valued functions of real
variables $x$ and $y$. If there exist continuous partial derivatives
at $z=z_0$ and they satisfy a differential constraint in the form of
coupled differential equations given by
\begin{equation}
\label{crc}
u_{,x}=v_{,y}\,;\ \ \ u_{,y}+v_{,x}=0,
\end{equation}
one can show that $f$ is complex differentiable, that is,
there exists a unique complex derivative
\begin{displaymath}
f^\prime(z_0)=\lim_{z\rightarrow z_0}\frac{f(z)-f(z_0)}{z-z_0}
\end{displaymath}
at $z=z_0$. (Here and in the following appendices, the subscripted comma
notation for partial derivatives is used, that is, $u_{,x}=
\upartial_{x}u=\partial u/\partial x$.) Furthermore, if the
condition given by equation~(\ref{crc}) is met for all points in $R$,
then the function $f$ is complex differentiable arbitrarily many
times (i.e., infinitely complex differentiable) at every point in $R$.
In general, a complex-valued complex-variable function $f$ is
referred to as an \emph{analytic} function (or a holomorphic
function) in $R$ if $f$ is complex
differentiable at every point in $R$. Note that $f$ is analytic in $R$
if and only if it satisfies the condition given by equation~(\ref{crc})
in $R$, which is usually referred to as \emph{Cauchy-Riemann
condition}. If $f=u+\I v$ is an analytic function, Cauchy-Riemann
condition~(\ref{crc}) implies that
\[
u_{,xx}+u_{,yy}=(u_{,x})_{,x}+(u_{,y})_{,y}=
(v_{,y})_{,x}+(-v_{,x})_{,y}=v_{,xy}-v_{,xy}=0,
\]
\[
v_{,xx}+v_{,yy}=(v_{,x})_{,x}+(v_{,y})_{,y}=
(-u_{,y})_{,x}+(u_{,x})_{,y}=-u_{,xy}+u_{,xy}=0.
\]
That is, both component functions $u$ and $v$ satisfy the
(two-dimensional) Laplace equation. In general, a function
$F:\mathbf R^2\rightarrow\mathbf R^2$ is referred to as a
\emph{harmonic} function if it satisfies Laplace equation
$\nabla^2F=(\upartial_x^2+\upartial_y^2)F=F_{,xx}+F_{,yy}=0$. In other
words, both the real and imaginary parts of any analytic function are
harmonic.

If one considers a complex conjugation $\z=x-\I y$ of a complex
variable, the function $f=u+\I v$ may be regarded as a function of
$z$ and $\z$, by the relations $x=(z+\z)/2$ and $y=\I(\z-z)/2$. Then,
formally, one may consider a partial derivative
\[
f_{,\z}=u_{,\z}+\I v_{,\z}=
(u_{,x}x_{,\z}+u_{,y}y_{,\z})+\I(v_{,x}x_{,\z}+v_{,y}y_{,\z})=
\frac{(u_{,x}+\I u_{,y})+\I(v_{,x}+\I v_{,y})}{2}=
\frac{u_{,x}-v_{,y}}{2}+\I\frac{u_{,y}+v_{,x}}{2}.
\]
That is, $f_{,\z}=0$ if and only if $f$ is analytic. In other words,
an analytic function may be casually understood as a complex
function of $z$ alone. (Roughly speaking, $f$ is analytic if it can be
expressed by complex constants and elementary differentiable real
functions with their arguments replaced by the complex variable $z$.)

If $f$ is analytic in $R$ containing $z_0$, there exists a convergent
Taylor series expression of $f$ at $z_0$
\begin{equation}
\label{tay}
f(z)=\sum_{n=0}^\infty(z-z_0)^n\frac{f^{(n)}(z_0)}{n!}=
f(z_0)+f^\prime(z_0)(z-z_0)+\frac{f^{\prime\!\prime}(z_0)}{2}(z-z_0)^2
+\frac{f^{\prime\!\prime\!\prime}(z_0)}{6}(z-z_0)^3+\cdots,
\end{equation}
where $f^{(n)}(z)=d^nf/dz^n$. A Taylor series (absolutely) converges
within its radius of convergence. That is, there exists a positive real
number $r$ such that the series in equation~(\ref{tay}) converges if
$|z-z_0|<r$. A complex function $f$ is said to have a \emph{pole} of
order $n$ at $z_0$ if $\lim_{z\rightarrow z_0}[(z-z_0)^mf(z)]$ is
unbounded for non-negative integers $m<n$ and $(z-z_0)^mf(z)$ is
analytic in a region containing $z_0$ for positive integers $m\ge n$.
If $z_0\in R$ is a pole of order $n$ of a function $f$ that is
analytic in $R-\{z_0\}$, then $f$ has a series expression of the
form of
\[
f(z)=\frac{(z-z_0)^nf(z)}{(z-z_0)^n}=
(z-z_0)^{-n}\sum_{m=0}^\infty\frac{1}{m!}
\left.\frac{d^m}{dz^m}\left[(z-z_0)^nf(z_0)\right]\right|_{z=z_0}
(z-z_0)^m=\sum_{m=-n}^\infty\frac{1}{(m+n)!}
\left.\frac{d^{m+n}}{dz^{m+n}}
\left[(z-z_0)^nf(z_0)\right]\right|_{z=z_0}(z-z_0)^{m},
\]
which converges if
$|z-z_0|<r$ and $z\ne z_0$ where $r$ is the radius of
convergence. Note that the series expansion of a complex function of
the form of $f=\sum_{m=-\infty}^\infty a_m(z-z_0)^m$ is generally
referred to as \emph{Laurent series}. In other words, if the $a_m$'s are
coefficients of the Laurent series expansion of $f$ at a pole of order
$n$, then $a_m=0$ for negative integers $m<-n$ and $a_{-n}\ne0$. If
$f$ is analytic in $\mathbf C$ except at isolated poles, the radius
of convergence for a Taylor series at a point $z_0$ that is not a pole is
the same as the distance to the nearest pole while that for a Laurent
series at a pole is also given by the distance to the next nearest
pole.

\section{lens equation with complex numbers}
\label{asec_ceq}

In general, the lens equation for a given lensing potential $\psi$ is
given by
\begin{equation}
\label{rle}
\bmath y=\bmath x-\bmath{\nabla_x}\psi(\bmath x),
\end{equation}
where $\bmath y$ is the angular position of the source and
$\bmath x$ is the angular position of the image. In addition, the
lensing potential is related to the mass column density \emph{per
unit cross sectional area} of lens $\Sigma$ via Poisson's equation
\begin{equation}
\label{rpe}
\nabla^2_{\bmath{x}}\psi=\frac{8\upi GD_\mathrm{R}}{c^2}\Sigma,
\end{equation}
where $D_\mathrm{R}\equiv D_\mathrm{LS}D_\mathrm{L}D_\mathrm{S}^{-1}$ is the
reduced distance\footnote{If $D_\mathrm{S}=D_\mathrm{L}+D_\mathrm{LS}$, then
$D_\mathrm{R}^{-1}=D_\mathrm{L}^{-1}+D_\mathrm{LS}^{-1}$.}. Note that one may
also rewrite equation~(\ref{rpe}) using the surface mass density
\emph{per unit solid angle} $\Sigma_A=\Sigma D_\mathrm{L}^2$
as in $\nabla^2_{\bmath{x}}
\psi=(8\upi GD_\mathrm{rel}^{-1}c^{-2})\Sigma_A$. Here, also used is that
$D_\mathrm{R}D_\mathrm{rel}=D_\mathrm{L}^2$.

By choosing some axis as a real axis, one may rewrite the lens
equation~(\ref{rle}) using complex numbers;
\begin{equation}
\label{cle}
\zeta=z-\alpha(z,\z).
\end{equation}
where $z=x_1+\I x_2$, $\zeta=y_1+\I y_2$ and $\alpha=\psi_{,1}+
\I\psi_{,2}$. Here $\psi_{,1}=\psi_{,x_1}$ and so on. The
complex deflection function $\alpha$ is related to the lensing
potential $\psi$ by
\[
\psi_{,\z}=\psi_{,1}x_{1,\z}+\psi_{,2}x_{2,\z}
=\frac{\psi_{,1}+\I\psi_{,2}}{2}=\frac{\alpha}{2},
\]
\begin{equation}
\label{ckv}
\alpha_{,z}=\alpha_{,1}x_{1,z}+\alpha_{,2}x_{2,z}
=\frac{(\psi_{,11}+\I\psi_{,21})-\I(\psi_{,12}+\I\psi_{,22})}{2}
=\frac{(\psi_{,11}+\psi_{,22})+\I(\psi_{,21}-\psi_{,12})}{2}
=\frac{1}{2}\nabla^2_{\bmath{x}}\psi=\frac{4\upi GD_\mathrm{R}}{c^2}\Sigma.
\end{equation}
Here, note that $x_1=(z+\z)/2$ and $x_2=i(\z-z)/2$. (The above equations
also imply that $2F_{,\z}=F_{,1}+\I F_{,2}=
(\bmath\nabla F)_1+\I(\bmath\nabla F)_2$
and $4F_{,z\z}=\nabla^2F$ for
any real valued function $F$.) Sometimes,
$\alpha_{,z}$ and $\alpha_{,\z}$ are also referred to as the
convergence $\kappa$ and the shear $\gamma$, respectively, of the
lensing potential. Note the differential
relation connecting the convergence and the shear,
$\kappa_{,\z}=\alpha_{,z\z}=\gamma_{,z}$. The convergence
$\kappa$ is always a non-negative real number, which is related to the
mass column density by $\kappa=(4\upi GD_\mathrm{R}c^{-2})\Sigma$ -- some
authors also define a critical mass column density $\Sigma_\mathrm{c}
\equiv c^2(4\upi GD_\mathrm{R})^{-1}$, and then $\kappa=
\Sigma/\Sigma_\mathrm{c}$. On the other hand, the shear $\gamma$ here is
a complex number (i.e.,it has two independent components);
\[
\gamma=\alpha_{,\z}=\alpha_{,1}x_{1,\z}+\alpha_{,2}x_{2,\z}
=\frac{(\psi_{,11}+\I\psi_{,21})+\I(\psi_{,12}+\I\psi_{,22})}{2}
=\frac{(\psi_{,11}-\psi_{,22})+\I(\psi_{,21}+\psi_{,12})}{2}
\]\[
|\gamma|^2=\alpha_{,\z}\overline{\alpha_{,\z}}
=\frac{1}{4}\left(\psi_{,11}-\psi_{,22}\right)^2+\psi_{,12}^2.
\]
It is notable, however, that the corresponding shear in real
2-dimensional notation does not behave as a vector under the
coordinate transformation (in particular, rotation of coordinate
axes). Rather it is a symmetric tensor of a rank of two with a null
contraction -- a real symmetric 2$\times$2 traceless matrix, that is,
\begin{equation}
\label{tsh}
\bmath\gammaup=\left(\begin{array}{cc}
\Re[\gamma]&\Im[\gamma]\\\Im[\gamma]&-\Re[\gamma]
\end{array}\right)=\frac{1}{2}\left(\begin{array}{cc}
\psi_{,11}-\psi_{,22}&2\psi_{,12}\\2\psi_{,21}&\psi_{,22}-\psi_{,11}
\end{array}\right)=|\gamma|\left(\begin{array}{cc}
\cos2\phi&\sin2\phi\\\sin2\phi&-\cos2\phi
\end{array}\right),
\end{equation}
where $2\phi$ is the argument of $\gamma$, i.e., $\gamma=|\gamma|
\E^{2\I\phi}$. The two eigenvalues of $\bmath\gammaup$ are
$\pm|\gamma|$ and its determinant is $-|\gamma|^2=-\gamma\bar\gamma$.
Note that the unit eigenvector associated with the positive eigenvalue
$|\gamma|$ is $(\cos\phi,\sin\phi)^T$ whose direction defines the
direction of the shear. On the other hand, the second unit eigenvector
$(-\sin\phi,\cos\phi)^T$ that is associated with the negative
eigenvalue $-|\gamma|$ is perpendicular to it. It is also notable that
the lensing shear field should satisfy the differential constraint that
$\gamma_{,zz}(=\kappa_{,z\z})\in\mathbf R$, that is,
$2{\gamma_\mathrm{R}}_{,12}={\gamma_\mathrm{I}}_{,11}-{\gamma_\mathrm{I}}_{,22}$,
where $\gamma_\mathrm{R}=\Re[\gamma]=|\gamma|\cos2\phi$ and
$\gamma_\mathrm{I}=\Im[\gamma]=|\gamma|\sin2\phi$. This constraint is
the direct result of the fact that the lensing deflection
can be described by a scalar potential -- i.e.,
$\gamma_{,zz}=\kappa_{,z\z}=4\nabla^2\kappa=4\nabla^2(2\nabla^2\psi)
=8\nabla^4\psi$ and $\nabla^4$ is a scalar operator.

In the case of the null convergence ($\kappa=0$), equation~(\ref{rpe})
becomes Laplace equation so that $\psi$ becomes a harmonic function.
Then, one can find a complex analytic function
$\psi_c(z)=\psi(x_1,x_2)+\I\varphi(x_1,x_2)$, where a function $\varphi$
satisfies the Cauchy-Riemann condition for $\psi_c(z)$. Then,
it is easy to show that
\[
\psi_c^\prime=\psi_{c,z}=\psi_{,z}+\I\varphi_{,z}=
\frac{\psi_{,1}-\I\psi_{,2}}{2}+\I\frac{\varphi_{,1}-\I\varphi_{,2}}{2}=
\frac{1}{2}\left(\psi_{,1}-\I\psi_{,2}-\I\psi_{,2}+\psi_{,1}\right)
=\bar\alpha,
\]
which also indicates that, for the case of the
null convergence, $\bar\alpha$ becomes also a complex analytic
function since it is a total complex derivative of an analytic
function.

If the lensing potential is given by
\[
\psi_c=\frac{4Gm}{D_\mathrm{rel}c^2}\ln(z-l_z);
\]\[
\psi=\frac{\psi_c+\bar\psi_c}{2}=
\frac{2Gm}{D_\mathrm{rel}c^2}\left[\ln(z-l_z)+\ln(\z-\lb_z)\right]=
\frac{2Gm}{D_\mathrm{rel}c^2}\ln[(z-l_z)(\z-\lb_z)]=
\frac{4Gm}{D_\mathrm{rel}c^2}\ln|z-l_z|, 
\]
the corresponding convergence is found to be
\[
\kappa=\frac{4\upi Gm}{D_\mathrm{rel}c^2}\delta^2(\bmath x-\bmath l)
=\frac{1}{\Sigma_\mathrm{c}}\frac{m}{D_\mathrm{rel}D_\mathrm{R}}
\delta^2(\bmath x-\bmath l),
\]
where $\delta^2(\bmath x)$ is the 2-dimensional Dirac delta function and
$\bmath l$ is the lens position in real 2-dimensional notation. In
other words, $\Sigma=mD_\mathrm{L}^{-2}\delta^2(\bmath x-\bmath l)$,
so that the potential describes lensing by
a point mass $m$ in empty space lying at $\bmath l$ (note that
$D_\mathrm{rel}D_\mathrm{R}=D_\mathrm{L}^2$).
Moreover, since Poisson's equation is linear in mass
(density), the lensing by a finite number of multiple point masses can
be described by the sum of individual potentials. Then, the deflection
function is found to be
\[
\alpha=\overline{\psi_c^\prime}=
\frac{4G}{D_\mathrm{rel}c^2}\sum_k\frac{m_k}{\z-\lb_{z_k}}.
\]
With this deflection function,
the lens equation~(\ref{cle}) is reduced to a particularly simple
(dimensionless) form
\begin{equation}
\label{mle}
\zeta=z-\sum_k\frac{q_k}{\z-\lb_{z_k}}\,;\ \ \
q_k=\frac{m_k}{M},
\end{equation}
provided that every angular measurement is
rescaled by the angular Einstein radius (eq.~\ref{enr})
corresponding to some mass $M$.
Here, $m_k$ is the mass of $k$-th component located at the complex
(angular) position $l_{z_k}$. While it is customary that the lens
equation~(\ref{mle}) is rescaled by the Einstein ring radius
corresponding either to the total mass of system ($\sum_k q_k = 1$) --
usually when several component masses are comparable -- or to the most
massive component ($q_1=1$) -- usually when the mass of the system is
dominated by a single component --, it is equally valid to choose
any mass $M$ as a unit of mass measurement as long as it is connected
to the unit of angular measurement $\theta$ by $D_\mathrm{rel}c^2\theta^2
=4GM$.

The local (differential) behaviour of the lens mapping can be studied
from the Jacobian matrix of the lens equation, which is basically the linear
transformation that approximates the lens mapping locally. For the
complex lens equation~(\ref{cle}), its Jacobian matrix is
\begin{equation}
\label{jcb}
\mathbfss J_z=\left(
\begin{array}{cc}
\zeta_{,z}&\zeta_{,\z}\\
\bar\zeta_{,z}&\bar\zeta_{,\z}
\end{array}\right)=\left(
\begin{array}{cc}
1-\alpha_{,z}&-\alpha_{,\z}\\
-\bar\alpha_{,z}&\overline{1-\alpha_{,z}}
\end{array}\right)=\left(
\begin{array}{cc}
1-\kappa&-\gamma\\-\bar\gamma&1-\kappa
\end{array}\right).
\end{equation}
Here, $\zeta_{,z}=1-\alpha_{,z}=1-\kappa\in\mathbf R$ and
$\zeta_{,\z}=-\alpha_{,\z}=-\gamma$. Since this matrix is Hermitian --
the corresponding Jacobian matrix of the lens equation~(\ref{rle})
with real 2-dimensional notation is real symmetric --, it has two real
eigenvalues, which are $\lambda_\pm=1-(\kappa\pm|\gamma|)$. Note that
$\lambda_++\lambda_-=2(1-\kappa)$ and $\lambda_+-\lambda_-=-2|\gamma|
\leq0$. Thus, $\lambda_+^2-\lambda_-^2=4|\gamma|(\kappa-1)$ and so
$|\lambda_+|<|\lambda_-|$ if $\kappa<1$ and vice versa. Suppose that
the principal argument of $\gamma=|\gamma|\E^{2\I\phi}$ is $2\phi\in
[0,2\upi)$. Then, one finds that
$E_+=\E^{\I\phi}$ and $E_-=\I E_+$ are
the unit eigenvectors of
Jacobian matrix, whose associated eigenvalues are $\lambda_\pm$,
respectively. Note that these two eigenvectors of the Jacobian matrix are also
the eigenvectors of the shear tensor (eq.~\ref{tsh}). This is a
natural consequence of the fact that $\mathbfss J=(1-\kappa)\mathbfss I-
\bmath\gammaup$, where $\mathbfss J$ is the real 2-dimensional Jacobian
matrix and $\mathbfss I$ is a 2$\times$2 identity matrix. Finally,
Jacobian determinant is invariant under the transformation of the lens
equation~(\ref{rle}) to the complex equation~(\ref{cle}), that is,
\[
\D\zeta\wedge\D\bar\zeta=(\D x_1+\I\D y_2)\wedge(\D y_1-\I\D y_2)
=-2\I\,\D y_1\wedge\D y_2=-2\I\J\,\D x_1\wedge\D x_2
=-2\I\J\,\left(\frac{\D z+\D\z}{2}\right)\wedge
\left(\frac{\D z-\D\z}{2\I}\right)=\J\,\D z\wedge\D\z.
\]
Hence,
\[
\J=\det\mathbfss J_z=\lambda_+\lambda_-=
(1-\kappa-|\gamma|)(1-\kappa+|\gamma|)=
(1-\kappa)^2-|\gamma|^2.
\]

\section{second order perturbation caustic}
\label{asec_spc}

\renewcommand{\thefigure}{\arabic{figure}}
\setcounter{figure}{3}
\begin{figure}
\plot{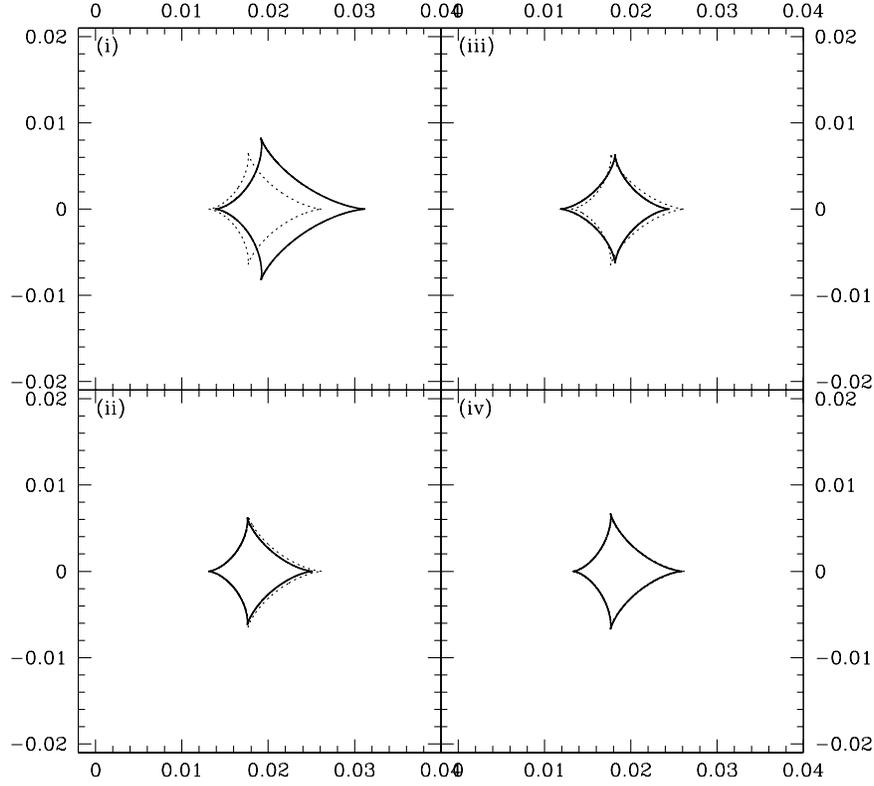}
\caption{
Various approximation of the central caustic of the system described
by the lens equation; $\zeta=z-\z^{-1}-q(\z-\z_\mathrm{p})^{-1}$ where $q=0.1$
and $z_\mathrm{p}=0.2$. The horizontal direction is same as the real direction.
The parameters are chosen such that the linear approximation of
Section~\ref{ppp} deviates noticeably from the true caustic although the
perturbative approach does not fail completely. The curves shown in
solid lines are (i) the linear order approximation in Section~\ref{ppp},
(ii) the second order approximation in Appendix~\ref{asec_spc}, (iii)
the linear caustic of the perturbation series of Section~\ref{ebp} keeping
only the quadrupole term, and (iv) the linear caustic of the
perturbation series of Section~\ref{ebp} keeping up to the octupole
term. Also shown in dotted lines are the `true' caustic found by the
method of \citet{Wi90}.
}\label{fca}
\end{figure}

With the lens mapping given by equation~(\ref{ple}), the linear
perturbation approach becomes inadequate or at least insufficient as
the magnitude of the perturbation grows. Here, the discussion in
Section~\ref{ssec_ccc} is extended to find the appropriate expression
of the caustic up to second order of the perturbation.

When the lens mapping is described by equation~(\ref{ple}), its
Jacobian determinant is found as equation~(\ref{pjd}) or 
\begin{equation}
\label{pj2}
\J=1-\frac{1}{z^2\z^2}
+\epsilon\left(\frac{f^{(1)}}{\z^2}+\frac{\f^{(1)}}{z^2}\right)
-\epsilon^2f^{(1)}\f^{(1)}.
\end{equation}
If one Taylor-expands
equation~(\ref{pj2}) at $z_0=\E^{i\phi}$ and keeps up to the second
order terms of $\delta z=z-z_0$ and $\epsilon$,
\begin{equation}
\label{tjd}
\J=
\epsilon(g_1+\g_1)-\epsilon^2g_1\g_1
+\left[2+\epsilon(g_2-2\g_1)\right]\E^{-\I\phi}\delta z
+\left[2+\epsilon(\g_2-2g_1)\right]\E^{\I\phi}\delta\z
-3\E^{-2\I\phi}\delta z^2-4\delta z\delta\z-3\E^{2\I\phi}\delta\z^2,
\end{equation}
where $g_n(\phi)=\E^{(n+1)\I\phi}f^{(n)}(\E^{\I\phi})$. Also used are
\[
\upartial_z\J=\frac{2}{z^3\z^2}
+\epsilon\left(\frac{f^{(2)}}{\z^2}-\frac{2\f^{(1)}}{z^3}\right)
+\mathcal O(\epsilon^2)
\,;\ \ \
\upartial_{\z}\J=\frac{2}{z^2\z^3}
+\epsilon\left(-\frac{2f^{(1)}}{\z^3}+\frac{\f^{(2)}}{z^2}\right)
+\mathcal O(\epsilon^2)
\]\[
\upartial_z^2\J=-\frac{6}{z^4\z^2}
+\mathcal O(\epsilon)
\,;\ \ \
\upartial_{\z}\upartial_z\J=-\frac{4}{z^3\z^3}
+\mathcal O(\epsilon)
\,;\ \ \
\upartial_{\z}^2\J=-\frac{6}{z^2\z^4}
+\mathcal O(\epsilon).
\]
Next, one inserts $\delta z=(\epsilon\chi_1+\epsilon^2\chi_2)\E^{\I\phi}$
into equation~(\ref{tjd});
\[
\J=
\epsilon(4\chi_1+g_1+\g_1)+\epsilon^2\left[
4\chi_2-g_1\g_1+(g_2+\g_2)\chi_1-2(g_1+\g_1)\chi_1-10\chi_1^2
\right]+\mathcal O(\epsilon^3).
\]
Here, $\chi_1$ and $\chi_2$ are assumed to be real-valued. To find the
expression for the second order approximation of the critical curve,
one solves for $\J=0+\mathcal O(\epsilon^3)$, that is,
\[
\chi_1(\phi)=-\frac{g_1+\g_1}{4}
\,;\ \ \
\chi_2(\phi)=\frac{2\chi_1^2-(g_2+\g_2)\chi_1+g_1\g_1}{4}=
\frac{(g_1+\g_1)^2}{32}
+\frac{(g_1+\g_1)(g_2+\g_2)}{16}+\frac{g_1\g_1}{4}.
\]
Then, the critical curve is found to be
$z_c(\phi)=z_0+\delta z=\E^{\I\phi}(1+\epsilon\chi_1+\epsilon^2\chi_2)$,
which, up to the linear approximation, recovers equation~(\ref{lca}).
The corresponding caustic can be found as the (second-order) image
of $z_c(\phi)$ under the lens mapping~(\ref{ple}). To do this, first,
one similarly Taylor-expands equation~(\ref{ple}) at $z_0=\E^{\I\phi}$
and keeps up to second order terms of $\epsilon$ and $\delta z$;
\[
\zeta=\delta z+\E^{2\I\phi}\delta\z-\E^{3\I\phi}\delta\z^2
-\epsilon\f_0-\epsilon\f_0^{(1)}\delta\z
=\E^{\I\phi}\left(\E^{-\I\phi}\delta z+\E^{\I\phi}\delta\z
-\E^{2\I\phi}\delta\z^2
-\epsilon\g_0-\epsilon\g_1\E^{\I\phi}\delta\z\right),
\]
and replaces $\delta z$ by $(\epsilon\chi_1+\epsilon^2\chi_2)
\E^{\I\phi}$ and takes up to the quadratic terms of $\epsilon$;
\begin{equation}
\zeta_c(\phi)=
\E^{\I\phi}\left[\epsilon(2\chi_1-\g_0)
+\epsilon^2(2\chi_2-\chi_1^2-\g_1\chi_1)\right],
\end{equation}
which is again reduced to equation~(\ref{lcu}) and consequently
equation~(\ref{cau}) if one keeps only up to the linear term. One
thing to note regarding the second order approximation, albeit
obvious, is that the expression is no longer linear in the
perturbation so that the superposition principle does not hold. That
is, to properly account for all the cross terms, one always needs to work
with all components of the perturbation together. An example of
the second order perturbation caustic is shown in Fig.~\ref{fca}
along with alternative approximations for the same underlying caustic.

\label{finish}

\end{document}